\documentclass[aps,twocolumn,showkeys,superscriptaddress,10pt]{revtex4-2}

\usepackage{amsfonts}
\usepackage{eurosym}
\usepackage{amssymb}
\usepackage[utf8]{inputenc}
\usepackage{color}
\usepackage{amsmath}
\usepackage{graphicx}
\usepackage[colorlinks={true}]{hyperref}
\hypersetup{colorlinks=true,linkcolor=red,citecolor=blue,urlcolor=blue}
\usepackage{dcolumn}
\usepackage{bm}
\usepackage{wrapfig}
\usepackage{subfigure}
\usepackage{ucs}
\usepackage{lineno}
\usepackage{epsfig}
\usepackage{psfrag}
\usepackage{epstopdf}
\usepackage{bigstrut}
\usepackage{multirow}
\usepackage{tikz}
\usepackage{orcidlink}
\usepackage{pstricks}
\usepackage{blkarray}
\newcommand{\matindex}[1]{\mbox{\scriptsize#1}}

\begin{document}

\title{Quantum Information Resources in Spin-1 Heisenberg Dimer Systems}

\author{Fadwa Benabdallah \orcidlink{0000-0003-1116-8741}}
\affiliation{LHEP-MS, Faculty of Sciences, Mohammed V University in Rabat, Rabat, Morocco}

\author{M. Y. Abd-Rabbou \orcidlink{0000-0003-3197-4724}}
\affiliation{School of Physical Sciences, University of Chinese Academy of Sciences, Yuquan Road 19A, Beijing 100049, China}
\affiliation{Mathematics Department, Faculty of Science, Al-Azhar University, Nasr City 11884, Cairo, Egypt}

\author{Mohammed Daoud \orcidlink{0000-0002-8494-2796}}
\affiliation{LPMS, Department of Physics, Faculty of Sciences, University Ibn Tofail, Kenitra, Morocco}
\affiliation{Abdus Salam International Centre for Theoretical Physics, Strada Costiera 11, I-34151 Trieste, Italy}

\author{Saeed Haddadi \orcidlink{0000-0002-1596-0763}}\email{haddadi@semnan.ac.ir}
\affiliation{Faculty of Physics, Semnan University, P.O. Box 35195-363, Semnan, Iran}

\begin{abstract}
We explore the quantum information resources within bipartite pure and mixed states of the quantum spin-1 Heisenberg dimer system, considering some interesting factors such as the $l_{1}$-norm of quantum coherence, relative coherence, negativity, and steering, influenced by the magnetic field and uniaxial single-ion anisotropy. Through a thorough investigation, we derive the system's density operator at thermal equilibrium and establish a mathematical framework for analyzing quantum resource metrics. Our results unveil the system's behavior at absolute zero temperature. We further observe temperature's role in transitioning the system towards classical states, impacting coherence, entanglement, and steering differently. Notably, we find that increasing the exchange anisotropy parameter can reinforce quantum correlations while adjusting the uniaxial single-ion anisotropy influences the system's quantumness, particularly when it is positive. Some recommendations to maximize quantum coherence, entanglement, and steering involve temperature reduction, increasing the exchange anisotropy parameter, and carefully managing the magnetic field and uniaxial single-ion anisotropy parameter, highlighting the intricate interplay between these factors in maintaining the system's quantum properties.
\end{abstract}

\maketitle

\section{Introduction}
Quantum spin systems are a class of promising solid-state and magnetic material resources that have garnered significant attention in the context of quantum information tasks and quantum computation, owing to their ability to create and distribute quantum correlations \cite{yao2012scalable,leuenberger2001quantum}. In recent years, a variety of spin systems have been extensively explored as potential qubits (spin-$\frac{1}{2}$) and qutrits (spin-$1$), including superconducting devices \cite
{devoret2013superconducting}, quantum dots \cite{PhysRevResearch.2.013062}, trapped atomic ions \cite{blatt2008entangled}, and both electronic and nuclear spin systems \cite{wolfowicz2021quantum}. To capture the quantum characteristics presented in these spin systems, various quantifiers have been employed. For instance, the Heisenberg dimer has been studied using Zeeman splitting \cite{PhysRevB.102.184419}, while the trimer projection has been explored in the context of entanglement in the ground
and thermal states of spin tetramers \cite{PhysRevA.72.022314}. Notably, spin chains \cite{Bonner1981}, mixed spin Ising-Heisenberg branched chains \cite%
{PhysRevB.102.064414}, triangular spin tubes \cite{sugimoto2023quasi}, Ising-Heisenberg Cairo pentagonal chains, tetrahedral chains \cite%
{arian2020spin}, antiferromagnetic Heisenberg trimerized chains \cite{PhysRevB.103.184415}, alternating-bond mixed diamond chains \cite{hida2020ground}, skewed and Creutz ladders \cite{PhysRevB.101.195110,PhysRevA.101.033607}, have all been investigated for their quantum properties. Moreover, in the domain of solid-state magnetic
material systems, various avenues exist for investigating external influences such as inhomogeneous Zeeman coupling \cite{PhysRevB.84.054451},
zero-field splitting \cite{kruk2001nuclear}, magnetic fields \cite{benabdallah2020quantum,benabdallah2021quantum}, dipole-dipole interaction
\cite{khalil2022robustness}, noise interactions \cite{PhysRevE.106.034122,PhysRevE.106.0341221}, as well as symmetric and asymmetric Dzyaloshinsky–Moriya (DM) interactions \cite{Yurischev2022}.

Quantum correlations, encompassing concepts such as entanglement, steering,
and coherence, unveil profound inter-linkages between individual quantum
entities \cite{RevModPhys.81.865,RevModPhys.89.041003, RevModPhys.92.015001,artur2022}%
. These correlations not only offer a deeper understanding of the quantum
world but also hold immense potential for groundbreaking advancements in
diverse applications \cite{nielsen2010quantum,asadali2024,asadali20242}. Entanglement, arguably the
most iconic embodiment of quantum correlations, was first elegantly
articulated by Schr\"{o}dinger in $1935$ \cite{schrodinger1935mathematical}.
It manifests as a non-classical correlation between particles, where the
fate of one becomes instantaneously intertwined with its distant
counterpart, defying classical notions of independence \cite%
{PhysRevLett.80.5239}. Its extensive exploration has yielded remarkable
applications in quantum communication \cite{PhysRevLett.91.207901} and
quantum computation \cite{mehring1999concepts}.
Beyond entanglement lies the
intriguing concept of steering, also introduced by Schr\"{o}dinger \cite%
{schrodinger1935mathematical}. It delves into scenarios where information
about one system influences the state of another, even without any direct
interaction \cite{obada2024}. Unlike entanglement, steering can occur even in non-local and
asymmetric scenarios \cite{PhysRevA.94.062120}. Its instrumental role in
cryptographic protocols \cite{PhysRevA.99.012302} and quantum teleportation
\cite{khalil2022robustness,PhysRevLett.70.1895,F1,F2} has drawn significant
attention, actuating some foundational and practical investigations \cite%
{PhysRevLett.125.020404}. Complementing these concepts is quantum coherence,
a fundamental principle stemming from the superposition principle \cite%
{RevModPhys.89.041003}. It embodies the ability of a quantum system to exist
in superposition states \cite{PhysRevA.98.052351}. Its applications range
from sophisticated quantum sensing to high-precision quantum metrology \cite%
{RevModPhys.89.035002,PhysRevA.98.032101}. Understanding these quantum correlations is imperative for unlocking the full potential of quantum systems. These three measures, tailored to capture different aspects of
these phenomena across various systems--from trapped ions \cite%
{ruster2016long} and solid states models \cite{F3} to atomic-field systems
\cite{abd2022enhancing} and quantum dots \cite{stockill2016quantum}--provide
valuable insights and pave the way for further exploration and innovation.

In the realm of spin-1 Heisenberg systems, the exploration of quantum correlations has been a focal point, considering different configurations such as bilinear and biquadratic interactions \cite{NN1,NN2}. Steering has
not been extensively studied in the context of high-dimensional spin-1
Heisenberg systems and coherence has been narrowly investigated compared to
entanglement, which has been widely discussed in many spin-1 Heisenberg
systems. The thermal coherence by $l_{1}$ norm and relative entropy of
spin-1 Heisenberg under diverse environmental conditions has been explored, including the XY model in the presence of an external magnetic field \cite{NNN2}. The thermal entanglement
properties of spin-1 Heisenberg interacted with inhomogeneous magnetic
fields \cite{NN3} and the simultaneous influences of both magnetic
field and DM interaction \cite{NN5} have been
investigated { (see also \cite{added1,added2,added3}).}

Hence, the motivation of this paper is to probe the thermal behavior of quantum information resources, such as coherence,
relative entropy of coherence, negativity, and entropic uncertainty-steering (EU-steering), in a bipartite spin-1 dimer quantum system under various intrinsic and extrinsic
parameters. These parameters include magnetic field, uniaxial single-ion
anisotropy, temperature, and exchange anisotropy.  Note that the homodinuclear nickel complex $\left[\mathrm{Ni}_2(\mathrm{Medpt})_2(\mu-\mathrm{ox})\left(\mathrm{H}_2 \mathrm{O}\right)_2\right]\left(\mathrm{ClO}_4\right)_2 \cdot 2 \mathrm{H}_2 \mathrm{O}$ provides an experimental realization of the antiferromagnetic spin-1 Heisenberg dimer \cite{Escuer1994,Ap1,Ap2}. This paper aims to understand how these parameters
affect the quantum resources within the system and whether certain parameter
combinations can enhance or diminish the characteristic traits of the
system. By analyzing the evolution of the different quantum quantifiers in
the parameter space, we aim to uncover patterns and trends that provide
insights into the thermal evolution of quantum correlations in the system.
Our findings suggest that certain parameter regimes lead to the enhancement
or suppression of quantum correlations, with implications for the system's
behavior and potential applications in quantum technologies. { By expanding our understanding of quantum coherence and correlations, this research may pave the way for unlocking the full potential of quantum technologies in various fields ranging from information processing to material science.}

The paper is organized in the following manner: Sec. \ref{SEC-2}
introduces a spin-1 Heisenberg model, in which we derive an exact solution
of the density matrix state. In Sec. \ref{SEC-3}, we introduce the quantum
information quantifiers. Sec. \ref{SEC-4} presents the thermal correlation
dependencies as functions of intrinsic and extrinsic parameters. Concluding
remarks and outlook are provided in Sec. \ref{Conclusions}.


\section{Model and Method}
\label{SEC-2}
The study of quantum spin systems has been a significant area of research in condensed matter physics due to its relevance in understanding various phenomena such as magnetism, superconductivity, and quantum phase transitions. Among these systems, the Heisenberg dimer, consisting of two interacting spins, serves as a fundamental model for investigating magnetic properties. In this theoretical exploration, we delve into the Hamiltonian description of the spin-1 Heisenberg dimer, elucidating its key components and their implications on the system's behavior. By considering the interplay of exchange interactions, magnetic anisotropy, and external fields, we aim to provide insights into the rich physical phenomena exhibited by this model system. Additionally, we discuss pertinent theoretical approaches employed in studying the spin-1 Heisenberg dimer, highlighting recent advancements and open questions in the field.
The comprehensive expression for the total effective Hamiltonian governing the spin-1 dimer is written as follows \cite{Ap1,Ap2}

\begin{equation}
\widehat{\mathcal{H}}=\widehat{\mathcal{H}}_{\text{exch}}+\widehat{\mathcal{H}}%
_{\text{azfs}}+\widehat{\mathcal{H}}_{\text{zeet}},  \label{H}
\end{equation}%
where the first part on the right-hand side of the Hamiltonian (\ref{H})
describes the exchange term, the second term is regarded as an axial zero-field splitting term, while the last part signifies the Zeeman term.
These terms are explicitly written as
\begin{eqnarray}
\widehat{\mathcal{H}}_{\text{exch}} =J_{x}S_{1}^{x}S_{2}^{x}+J_{y}S_{1}^{y}S_{2}^{y} +\Delta S_{1}^{z}S_{2}^{z},
\end{eqnarray}
\begin{eqnarray}
\widehat{\mathcal{H}}_{\text{azfs}} =D[(S_{1}^{z})^{2}+(S_{2}^{z})^{2}],
\end{eqnarray}
and
\begin{eqnarray}
\widehat{\mathcal{H}}_{\text{zeet}}=-h(S_{1}^{z}+S_{2}^{z}),
\end{eqnarray}
where $S_{i}^{\alpha }\left( \alpha =x,y,z\text{ with }i=1,\text{ }2\right) $
denotes the three spatial components of the spin-1 operator associated with
two distinct magnetic ions $($we set $\hbar =1)$. { The coupling coefficient $J_{x}=J_{y}=J$ specifies the magnitude of the XXZ exchange interaction. Here, let us consider $\Delta$ as the \textit{easy-axis exchange anisotropy} which takes
the real values.}
Moreover, $D$
stands for the uniaxial single-ion anisotropy, and the term $h=g\mu _{B}B$
accommodates Zeeman's effect resulting from the external magnetic field $B$,
with $\mu _{B}$ representing the Bohr magneton and $g$ being the Land\'{e} $%
g $-factor. The energy eigenvalues of the
spin-1 Heisenberg dimer and the associated eigenstates, as defined by Hamiltonian (\ref{H}), are explicitly
reported in Appendix \ref{Sec:appendix a}.

To incorporate the thermal interaction into the spin-1 Heisenberg dimer system, the thermal state of the system, maintained in equilibrium with a
thermal reservoir at temperature $T$, is described by the Gibbs density
operator as
\begin{equation}
\rho (T)=\frac{1}{\mathcal{Z}}\exp ( -\beta \widehat{\mathcal{H}}%
) =\frac{1}{\mathcal{Z}}\sum_{i=1}^{9}\exp \left( -\beta E_{i}\right)
\left\vert \psi _{i}\right\rangle \left\langle \psi _{i}\right\vert ,
\end{equation}%
where $\mathcal{Z}=\sum_{i=1}^{9}\exp \left( -\beta E_{i}\right) $ is the
partition function, $\beta ={1%
\mathord{\left/
        {\vphantom {1 {{k_B}T}}} \right.\kern-\nulldelimiterspace} {{k_B}T}}$
in which $k_{B}$ is the Boltzmann constant while $T$ is the absolute
temperature. Therefore, the state of the system at thermal equilibrium can
be written, in the computational basis $\{\left\vert i,j\right\rangle ;%
\hspace{0.1cm}i=-1,0,1\hspace{0.1cm}\mathrm{and}\hspace{0.1cm}$ $j=-1,0,1\}$%
, as follows
{ \begin{widetext}
\begin{eqnarray}\label{gho}
  \rho (T)=\mathcal{Z}^{-1} \times
  \begin{blockarray}{cccccccccc}
    & \matindex{$\vert -1,-1\rangle$} & \matindex{$\vert -1,0\rangle$} & \matindex{$\vert -1,1\rangle$} & \matindex{$\vert 0,-1\rangle$} & \matindex{$\vert 0,0\rangle$} & \matindex{$\vert 0,1\rangle$} & \matindex{$\vert 1,-1\rangle$} & \matindex{$\vert 1,0\rangle$} & \matindex{$\vert 1,1\rangle$}  \\
    \begin{block}{c(ccccccccc)}
      \matindex{$\langle -1,-1\vert$} & \rho _{0,0} & 0 & 0 & 0 & 0 & 0 & 0 & 0 & 0 \\
\matindex{$\langle -1,0\vert$} & 0 & \rho _{1,1} & 0 & \rho _{1,3} & 0 & 0 & 0 & 0 & 0 \\
\matindex{$\langle -1,1\vert$} & 0 & 0 & \rho _{2,2} & 0 & \rho _{2,4} & 0 & \rho _{2,6} & 0 & 0 \\
\matindex{$\langle 0,-1\vert$} & 0 & \rho _{1,3} & 0 & \rho _{1,1} & 0 & 0 & 0 & 0 & 0 \\
\matindex{$\langle 0,0\vert$} & 0 & 0 & \rho _{2,4} & 0 & \rho _{4,4} & 0 & \rho _{2,4} & 0 & 0 \\
\matindex{$\langle 0,1\vert$} & 0 & 0 & 0 & 0 & 0 & \rho _{5,5} & 0 & \rho _{5,7} & 0 \\
\matindex{$\langle 1,-1\vert$} & 0 & 0 & \rho _{2,6} & 0 & \rho _{2,4} & 0 & \rho _{2,2} & 0 & 0 \\
\matindex{$\langle 1,0\vert$} & 0 & 0 & 0 & 0 & 0 & \rho _{5,7} & 0 & \rho _{5,5} & 0 \\
\matindex{$\langle 1,1\vert$} & 0 & 0 & 0 & 0 & 0 & 0 & 0 & 0 & \rho _{8,8} \\
    \end{block}
  \end{blockarray}
\end{eqnarray}
\end{widetext}}
where the nonzero parameters are
{ \begin{equation*}
\rho _{0,0}=e^{-\beta E_{1}},
\text{ \ \ }\rho_{8,8}=e^{-\beta E_{4}},
\end{equation*}
\begin{equation*}
\rho _{1,1}=e^{-\beta(D-h)}\cosh{(\beta J)},
\end{equation*}
\begin{equation*}
\rho _{1,3}=-e^{-\beta(D-h)}\sinh{(\beta J)},
\end{equation*}
\begin{equation*}
\rho _{2,2}=\frac{e^{-\beta E_{5}}}{2+\Lambda _{+}^{2}}+
\frac{e^{-\beta E_{6}}}{2+\Lambda_{-}^{2}}
+\frac{e^{-\beta E_{7}}}{2},
\end{equation*}
\begin{equation*}
\rho _{2,4}=\Lambda_{+}\frac{e^{-\beta E_{5}}}{2+\Lambda _{+}^{2}} +\Lambda _{-}\frac{e^{-\beta E_{6}}}{2+\Lambda _{-}^{2}},
\end{equation*}
\begin{equation*}
\rho _{2,6}=\frac{e^{-\beta E_{5}}}{2+\Lambda _{+}^{2}}+
\frac{e^{-\beta E_{6}}}{2+\Lambda_{-}^{2}}
-\frac{e^{-\beta E_{7}}}{2},
\end{equation*}
\begin{equation*}
\rho _{4,4}=\Lambda _{+}^{2} \frac{e^{-\beta E_{5}}}{2+\Lambda
_{+}^{2}}+ \Lambda_{-}^{2} \frac{e^{-\beta E_{6}}}{2+\Lambda
_{-}^{2}},
\end{equation*}
\begin{equation*}
\rho _{5,5}=e^{-\beta (D+h)} \cosh{(\beta J)},
\end{equation*}
\begin{equation}
\rho _{5,7}=-e^{-\beta(D+h)}\sinh{(\beta J)}.
\end{equation}}
The partition function $\mathcal{Z}$ in \eqref{gho} is given by
{ \begin{align}
\mathcal{Z}= & e^{\beta(\Delta-2 D)}+2 e^{-\beta(\Delta+2 D)} \cosh (2 \beta h )\nonumber\\
&+2 e^{-\beta D}\left\{\cosh (\beta[h+J])
+\cosh (\beta[h-J])\right\}\nonumber\\
&+2 e^{\beta(\Delta-2 D)/2} \cosh (\beta \sqrt{(\Delta-2 D)^2+8 J^2} /2).
\end{align}}

\section{Quantum information quantifiers}
\label{SEC-3}

To characterize the quantum information resources, specifically quantum
steering, entanglement, and quantum coherence of the spin-1 Heisenberg dimer (%
\ref{H}), various metrics can be employed. This section briefly introduces
the mathematical forms of $l_{1}$-norm of coherence, relative entropy of
coherence, negativity, and EU-steering.

\subsection{Coherence measures}
Let's consider two readily computable coherence estimators to measure quantum coherence in the spin-1 Heisenberg system, namely the relative entropy of
coherence and the $l_{1}$-norm of coherence \cite{Baumgratz2014,Hu2018,COH}.
Using both of these measures in a given scenario allows for a comprehensive analysis of quantum coherence.

The relative
entropy of coherence $\mathcal{C}_{r}(\rho )$ is a distance-based coherence
quantifier that measures coherence using the minimal distance between the
quantum state and the set of incoherent states. Using the properties of
relative entropy, one can show that $\mathcal{C}_{r}(\rho )$ is expressed as

\begin{equation}
\mathcal{C}_{r}(\rho )=S(\rho _{\text{diag}})-S(\rho ),
\end{equation}%
where $\rho _{\text{diag}}$ describes the quantum incoherent state. { This
state is obtained by eliminating all off-diagonal
elements of $\rho $ under the reference basis $\{|i\rangle \}_{i=1}^{d}$ of
a $d$-dimensional Hilbert space $\mathcal{H}_{d}$. In this basis, $\rho _{
\text{diag}}$ is the incoherent diagonal matrix $\rho _{\text{diag}%
}=\sum_{i=1}^{d}\rho _{ii}|i\rangle \langle i|$.}

Thus, the relative entropy of coherence for our thermal state \eqref{gho} can be
obtained in the following form

\begin{equation}
\mathcal{C}_{r}[\rho(T)]=\sum_{k=1}^{9}\mu _{k}\log _{2}\mu
_{k}-\sum_{k=1}^{9}v_{k}\log _{2}v_{k},  \label{MIDSPIN}
\end{equation}%
where $\mu _{k}$ are the eigenvalues of the state $\rho (T)$,
given by%
\begin{align}
&\mu _{1;~2} =\frac{1}{\mathcal{Z}}(\rho _{1,1}\pm \rho _{1,3}),\text{ \ \ \
\ }\mu _{3;~4}=\frac{1}{\mathcal{Z}}(\rho _{5,5}\pm \rho _{5,7}), \nonumber\\
&\mu _{5} =\frac{1}{\mathcal{Z}}\rho _{0,0},\text{ \ }\mu _{6}=\frac{1}{
\mathcal{Z}}(\rho _{2,2}-\rho _{2,6}),\text{ \ \ }\mu _{7}=\frac{1}{\mathcal{Z}}\rho _{8,8}, \nonumber
\end{align}
and
\begin{align}
\mu _{8;~9} =\frac{1}{2\mathcal{Z}}&\bigg[\rho _{2,2}+\rho _{2,6}+\rho _{4,4}\nonumber\\
&\pm
\sqrt{8\rho _{2,4}^{2}+(\rho_{2,2}+\rho_{2,6}-\rho_{4,4})^2}\bigg],
\end{align}
besides, $v_{k}$ are the non-vanishing eigenvalues of the diagonal density matrix $
\rho _{\text{diag}}(T)$.

On the other hand, we employ the intuitive $l _{1}$-norm coherence
measure, which is defined as the sum of the absolute off-diagonal elements
of a quantum state $\rho $ in the reference basis $\{|i\rangle \}$. This
measure can be computed as

\begin{equation}
\mathcal{C}_{l _{1}}(\rho )=\sum_{i\neq j}\left\vert \rho
_{i,j}\right\vert .
\end{equation}%
Consequently, the corresponding $l _{1}$-norm of quantum coherence for
our system, as described by the thermal state $\rho (T)$ (\ref{gho}%
), is expressed as follows

\begin{equation}
\mathcal{C}_{l _{1}}[\rho(T)]=\frac{1}{\mathcal{Z}}\left( 2\left\vert \rho
_{1,3}\right\vert +4\left\vert \rho _{2,4}\right\vert +2\left\vert \rho
_{2,6}\right\vert +2\left\vert \rho _{5,7}\right\vert \right).
\end{equation}

\subsection{Negativity}

The concept of negativity is regarded as the most general quantity that can be straightforwardly used as a metric of the bipartite entanglement even for a spin-1 system. According to the Peres--Horodecki separability
criterion \cite{Peres,RevModPhys.81.865},
the negativity is related to the trace norm of $\rho ^{t_{A}}$ defined as \cite{Vidal}

\begin{equation}
\mathcal{N}(\rho )=\frac{\Vert \rho ^{t_{A}}\Vert _{1}-1}{2},
\end{equation}%
where trace norm is defined by $\Vert \rho ^{t_{A}}\Vert _{1}=\text{Tr}
\sqrt{(\rho ^{t_{A}})^{\dagger }\rho ^{t_{A}}}$. {
Some notes on negativity are provided in Appendix \ref{Sec:appendix b}.}

To compute the analytical form of negativity based on our thermal state $\rho (T)$, one needs to determine the
partially transposed density matrix $\rho^{t_{A}}(T)$.
{ In the same computational basis, $\rho ^{t_{A}}(T)$ takes the following form
\begin{eqnarray}\label{rhotransp}
&&\rho^{t_A} (T)=\nonumber\\
&&\frac{1}{\mathcal{Z}}\left(
	\begin{array}{ccccccccc}
		\rho _{0,0} & 0 & 0 & 0 & \rho _{1,3} & 0 & 0 & 0 & \rho _{2,6} \\
		0 & \rho _{1,1} & 0 & 0 & 0 & \rho _{2,4} & 0 & 0 & 0 \\
		0 & 0 & \rho _{2,2} & 0 & 0 & 0 & 0 & 0 & 0 \\
		0 & 0 & 0 & \rho _{1,1} & 0 & 0 & 0 & \rho _{2,4} & 0 \\
		\rho _{1,3} & 0 & 0 & 0 & \rho _{4,4} & 0 & 0 & 0 & \rho _{5,7} \\
		0 & \rho _{2,4} & 0 & 0 & 0 & \rho _{5,5} & 0 & 0 & 0 \\
		0 & 0 & 0 & 0 & 0 & 0 & \rho _{2,2} & 0 & 0 \\
		0 & 0 & 0 & \rho _{2,4} & 0 & 0 & 0 & \rho _{5,5} & 0 \\
		\rho _{2,6} & 0 & 0 & 0 & \rho _{5,7} & 0 & 0 & 0 & \rho _{8,8}%
	\end{array}%
	\right). \nonumber\\
\end{eqnarray}
Having the partially transposed matrix \eqref{rhotransp}, the negativity
of the state \eqref{gho} can be expressed as
\begin{eqnarray}
\mathcal{N}[\rho(T)] =\frac{1}{2\mathcal{Z}}&&\Big[\sqrt{\rho^{2} _{0,0}+\rho^{2}
_{1,3}+\rho^{2}_{2,6}}+2\sqrt{\rho^{2} _{1,1}+\rho^{2} _{2,4}}\notag \\
&&+2\sqrt{\rho^{2} _{2,4}+\rho^{2} _{5,5}}
+\sqrt{\rho^{2} _{1,3}+\rho^{2} _{4,4}+\rho^{2} _{5,7}}
\notag \\
&&+2\sqrt{
\rho^{2} _{2,2}}+\sqrt{\rho^{2} _{2,6}+\rho^{2}
_{5,7}+\rho^{2} _{8,8}}-\mathcal{Z}\Big].\notag \\
\end{eqnarray}}

\subsection{Entropic Uncertainty-Steering (EU-steering)}

Entropic uncertainty relations (EURs) are inequalities that place limits on
the amount of information that can be simultaneously known about a quantum
system. These inequalities can be used to derive steering inequalities,
which are tests of whether a quantum state can be explained by the local hidden
variables (LHVs). For discrete qudit quantum state $\rho$ with $N$ mutually
unbiased bases, the EUR in terms of the discrete Shannon entropy reads \cite%
{PhysRevA.79.022104}
{ \begin{equation}
\sum_{m=1}^{N} H(p_{m_i}) \geq N \log_2 \left(\frac{d \ N}{d \ \text{Tr} \rho^2+N-1}\right),
\end{equation}}
with $H(p_{m_i})=\sum_{i=1}^{d} p_{m_i}\log_2 \frac{1}{p_{m_i}}$, where  $p_{m_i}$ is the
probability of obtaining $m^{th}$ observable with dimension $d$ and the system is measured in the $i^{th}$ basis.

In EU-steering, the steering inequality is derived from EUR using the
discrete conditional Shannon entropy. For $2 \otimes 2 $ and $2\otimes3$%
-dimensional quantum states { in Hilbert space}, the EU-steering is investigated using Shannon
entropy \cite{S1,S2,S3}. This approach has several advantages over
traditional steering criteria. First, it can be used to detect weakly
steerable states, which are states that cannot be explained by LHVs with
high confidence. Second, it can be used to derive steering inequalities that
are tight in a variety of scenarios. The EU-steering between bipartite $A$
and $B$ states is given by \cite{costa2018entropic}
\begin{equation}
\sum_{m=1}^{N}H(p_{m_{i}}^{B}|p_{m_{j}}^{A})\geq \mathcal{B}_{A},
\end{equation}%
where $%
H(p_{m_{i}}^{B}|p_{m_{j}}^{A})=H(p_{m_{i},m_{j}}^{AB})-H(p_{m_{j}}^{A})$ is
the conditional Shannon entropy. However, the right-hand side is obtained
for $d\geq 2$ as \cite{PhysRevA.79.022104,costa2018entropic}
\begin{equation}
\mathcal{B}_{A}=N\log _{2}F+\bigg(N-F\frac{d_{A}+N-1}{d_{A}}\bigg)\log _{2}%
\bigg(1+\frac{1}{F}\bigg)^{1+F},
\end{equation}%
where $F$ is the floor function of $\frac{N\ d_{A}}{d_{A}+N-1}$. On the other
hand, for a qutrit system ($d=3$) and considering the three spin-1 components
are measurements, we have
\begin{equation}
H(\hat{S}_{x}^{B}|\hat{S}_{x}^{A})+H(\hat{S}_{y}^{B}|\hat{S}_{y}^{A})+H(\hat{%
S}_{z}^{B}|\hat{S}_{z}^{A})\geq \frac{8}{3},
\end{equation}%
where $H(\hat{S}_{m}^{B}|\hat{S}_{m}^{A})=-%
\sum_{i,j}^{3}p_{m_{i},m_{j}}^{AB}\log
_{2}p_{m_{i},m_{j}}^{AB}+\sum_{j}^{3}p_{m_{j}}^{A}\log _{2}p_{m_{j}}^{A}$
with $p_{m_{i},m_{j}}^{AB}=\langle m_{i},m_{j}|\rho _{AB}|m_{i},m_{j}\rangle
$, $p_{m_{j}}^{A}=\langle m_{j}|\rho _{A}|m_{j}\rangle $, and $%
\{m_{i},m_{j}\}$ are the set of eigenbasis of spin-1 components. Employing
the thermal density matrix \eqref{gho} and violating the previous inequality to
coincide with the other quantifiers, one can get the steering inequality as
\begin{widetext}
\begin{equation}\label{eusteering}
\mathcal{S}= \sum_{i,j=1}^{3}\big(p_{x_{i},x_{j}}^{AB}\log
_{2}p_{x_{i},x_{j}}^{AB}+p_{y_{i},y_{j}}^{AB}\log
_{2}p_{y_{i},y_{j}}^{AB}+p_{z_{i},z_{j}}^{AB}\log _{2}32\sqrt[3]{2}%
p_{z_{i},z_{j}}^{AB}\big) \\
 -\sum_{j=1}^{3}\big(p_{x_{j}}^{A}\log _{2}p_{x_{j}}^{A}+p_{y_{j}}^{A}\log
_{2}p_{y_{j}}^{A}+p_{z_{j}}^{A}\log _{2}p_{z_{j}}^{A}\big)\leq \frac{8}{3}
\end{equation}%
\end{widetext}
here
\begin{equation}
\begin{split}
& p_{x_{1},x_{1}}^{AB}
=p_{y_{1},y_{1}}^{AB}=p_{x_{2},x_{2}}^{AB}=p_{y_{2},y_{2}}^{AB}=\frac{1}{16}\big\{1+\zeta +\eta +\xi \big\},\\
&p_{x_{3},x_{3}}^{AB}=p_{y_{3},y_{3}}^{AB}=\frac{1}{4}\big\{\kappa +\xi \big\}, \\
&p_{x_{1},x_{2}}^{AB}
=p_{y_{1},y_{2}}^{AB}=p_{x_{2},x_{1}}^{AB}=p_{y_{2},y_{1}}^{AB}=\frac{1}{%
16}\big\{1+\zeta -\eta +\xi \big\},\\
& p_{x_{1},x_{3}}^{AB}=p_{y_{1},y_{3}}^{AB}=p_{x_{2},x_{3}}^{AB}=p_{y_{2},y_{3}}^{AB}\\
 &\; \; \quad \quad=\frac{1}{8}\big\{\kappa -\zeta +2(\rho _{3,3}+\rho
_{5,5})\big\},
\\
& p_{x_{3},x_{1}}^{AB}=p_{y_{3},y_{1}}^{AB}=p_{x_{3},x_{2}}^{AB}=p_{y_{3},y_{2}}^{AB}\\
 &\; \; \quad \quad=\frac{1}{8}\big\{\kappa -\zeta +2(\rho _{1,1}+\rho
_{7,7})\big\},
\\
& p_{z_{1},z_{1}}^{AB}=\rho _{8,8}, \quad \quad p_{z_{2},z_{2}}^{AB}=\rho _{4,4}, \quad \quad p_{z_{3},z_{3}}^{AB}=\rho _{0,0}, \\ &p_{z_{1},z_{2}}^{AB}=\rho _{7,7}, \quad \quad p_{z_{2},z_{1}}^{AB}=\rho _{5,5}, \quad \quad
 p_{z_{1},z_{3}}^{AB}=\rho _{6,6}, \\
&p_{z_{3},z_{1}}^{AB}=\rho _{2,2}, \quad \quad p_{z_{2},z_{3}}^{AB}=\rho _{3,3}, \quad \quad
p_{z_{3},z_{2}}^{AB}=\rho _{1,1},\\
& p_{x_{1}}^{A}=p_{x_{2}}^{A}=p_{y_{1}}^{A}=p_{y_{2}}^{A}=\frac{1}{4}\big\{1+(\rho _{3,3}+\rho _{4,4}+\rho _{5,5})%
\big\},\\
&p_{x_{3}}^{A}=p_{y_{3}}^{A}=
\frac{1}{2}\big\{1-(\rho _{3,3}+\rho _{4,4}+\rho _{5,5})\big\},
\\
& p_{z_{1}}^{A}=\rho _{0,0}+\rho _{1,1}+\rho _{2,2},\quad \quad
p_{z_{2}}^{A}=\rho _{3,3}+\rho _{4,4}+\rho _{5,5},\\
&p_{z_{3}}^{A}=\rho _{6,6}+\rho _{7,7}+\rho _{8,8},
\end{split}%
\end{equation}
with
\begin{equation}
\begin{split}
& \kappa =\rho _{0,0}+\rho _{2,2}+\rho _{6,6}+\rho _{8,8},\\ &\zeta =\rho
_{1,1}+\rho _{3,3}+3\rho _{4,4}+\rho _{5,5}+\rho _{7,7}, \\
& \eta =4\Re \lbrack \rho _{1,3}+\rho _{2,4}+\rho _{4,6}+\rho _{5,7}],\\
& \xi =2\Re \lbrack \rho _{2,6}].
\end{split}%
\end{equation}

\section{The behavior of quantum information resources}
\label{SEC-4}

\begin{figure*}[!t]
\centering
\includegraphics[scale=0.39,trim=00 00 00 00, clip]{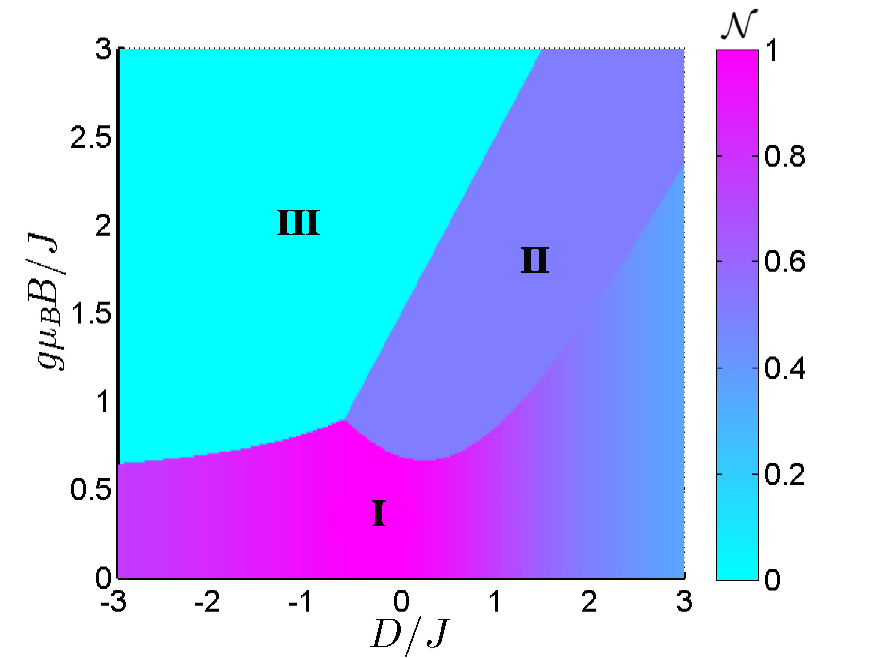} %
\includegraphics[scale=0.39,trim=00 00 00 00, clip]{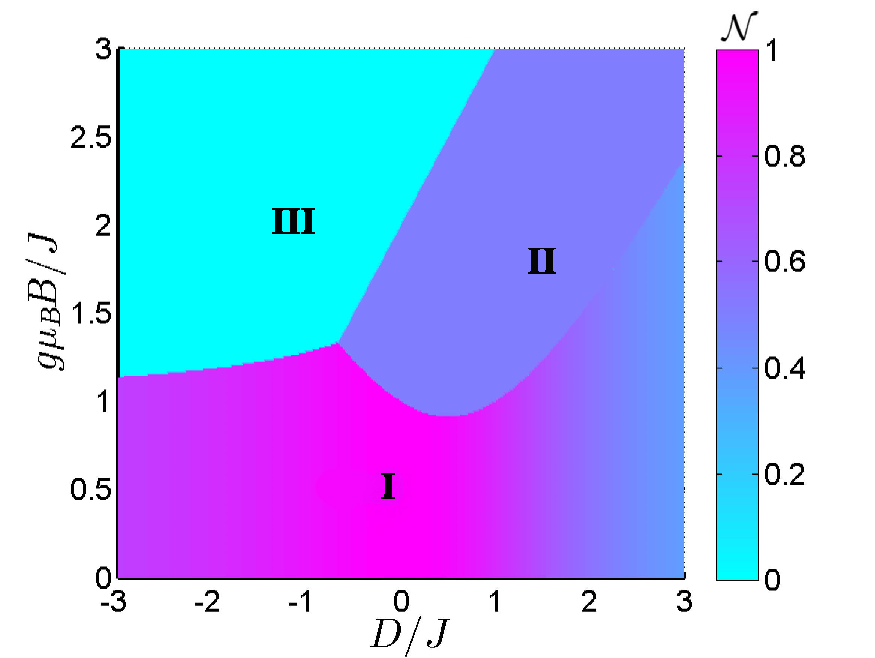} %
\includegraphics[scale=0.39,trim=00 00 00 00, clip]{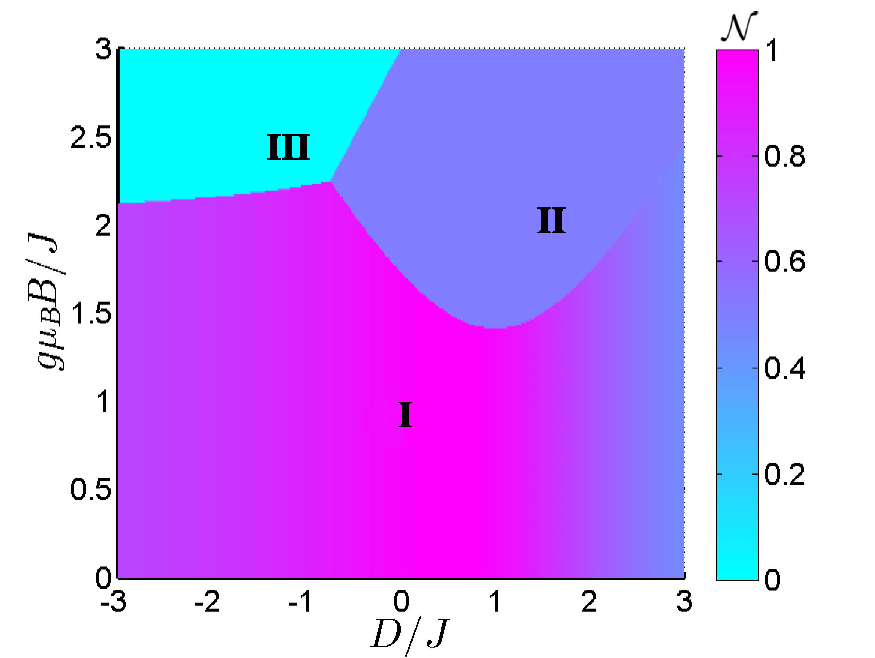}
\put(-477,107){$\textbf{(a)}$}\put(-313,107){ $\textbf{(b)}$}\put(-145,107){ $%
\textbf{(c)}$}
\caption{ Zero-temperature density plot of the negativity in the $
g\mu_{B}B/J-D/J$ plane for three different values of the exchange anisotropy: $%
\Delta/J = 0.5$ (a), $\Delta/J=1.0$ (b), and $\Delta/J=2.0$ (c).}
\label{Fig:1}
\end{figure*}

Before delving into a detailed examination of quantum resources, we undertake an in-depth examination of bipartite quantum
entanglement within the spin-1 Heisenberg dimer system at absolute zero
temperature, exploring three distinct settings for the relative strength of
exchange anisotropy $\Delta$. To accomplish this, we present the density plot of
negativity at zero temperature in Fig. \ref{Fig:1}, illustrating the uniaxial
single-ion anisotropy versus magnetic field plane for three representative
values of the easy-axis exchange anisotropy. These values encompass the fully isotropic
case with $\Delta/J =1.0$, and specific scenarios with the exchange anisotropies $\Delta/J
=0.5$ and $\Delta/J =2.0$. It should be noted that the ground states of a spin-1 Heisenberg dimer system have already been examined in Ref. \cite{Ap1}, and to avoid repetition, we will not discuss this issue.

The entanglement in terms of negativity
achieves its peak at $\mathcal{N}=1$ for specific combinations of exchange
and uniaxial single-ion anisotropies, indicating a maximally entangled state (see Region I).
As a significant result, negativity attains a value equal to half of its maximum,
$\mathcal{N}=0.5$. In this case, the system exhibits partial entanglement (see Region II).
In contrast, negativity tends to zero, emphasizing the distinct nature of
entanglement in a classical regime; a separable state (see Region III).

Notably, negativity undergoes a remarkable discontinuous jump at each boundary, separating
the regions I, II, and III.
This discontinuity underscores the dramatic shifts in entanglement characteristics as the system transitions between these regions. Indeed, our findings reveal that the system shows a maximally entangled state at absolute zero temperature for weak uniaxial single-ion anisotropy and magnetic field values. Increasing the magnetic field transitions the system to a separable state while increasing the uniaxial single-ion anisotropy leads to a partially entangled state. Also, enhancing the exchange anisotropy parameter significantly reinforces quantum correlations and thus increases the range of the entangled state. In summary, the interplay of quantum entanglement within different states reveals the richness and complexity of quantum systems, providing profound insights into the unique behaviors of entangled particles.
Moreover, by considering the
easy-axis exchange anisotropy $\Delta/J =0.5$ [Fig. \ref{Fig:1}(a)], we
observe a shift in the maximum of negativity $\mathcal{N}=1$ towards
the nonpositive values of single-ion anisotropy ($D/J<0$). Conversely, easy-axis exchange
anisotropy (e.g. $\Delta/J =2.0$) shifts this maximum towards
the positive single-ion anisotropy ($D/J>0$). Consequently, we conclude that
the spin-1 Heisenberg dimer exhibits the strongest quantum
entanglement, namely the highest negativity $\mathcal{N}=1$, either for the
fully isotropic case with $\Delta/J =1.0$ under the assumption that the
uniaxial single-ion anisotropy is absent $D/J=0$, i.e., it does not possess
any form of magnetic anisotropy, either when the easy-axis exchange anisotropy compensates for the effect of single-ion anisotropy, or vice versa.

In the cases where $\Delta/J =0.5$ and $\Delta/J =1.0$, as observed in Fig. \ref{Fig:1}(a) and Fig. \ref{Fig:1}(b), { separable states}
predominate over entangled ones, suggesting they are suboptimal for
maintaining and preserving the quantum property of the system over extended
interaction times. However, when the exchange anisotropy parameter takes $%
\Delta/J =2.0$ [see Fig. \ref{Fig:1}(c)], we observe the strongest and largest
amount of bipartite quantum entanglement compared to other scenarios.
Henceforth, our focus will be
on the case where the exchange interaction is $\Delta/J =2.0$, delving deeper into
the physical implications of this configuration.

\begin{figure*}[!t]
\centering
\includegraphics[scale=0.48,trim=00 00 00 00, clip]{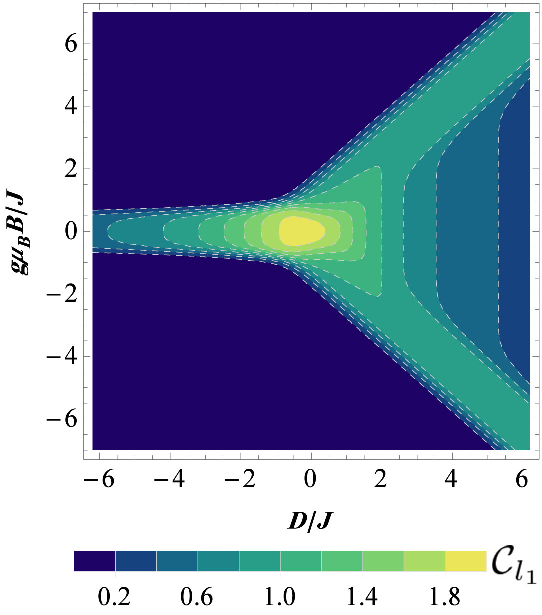} %
\includegraphics[scale=0.48,trim=00 00 00 00, clip]{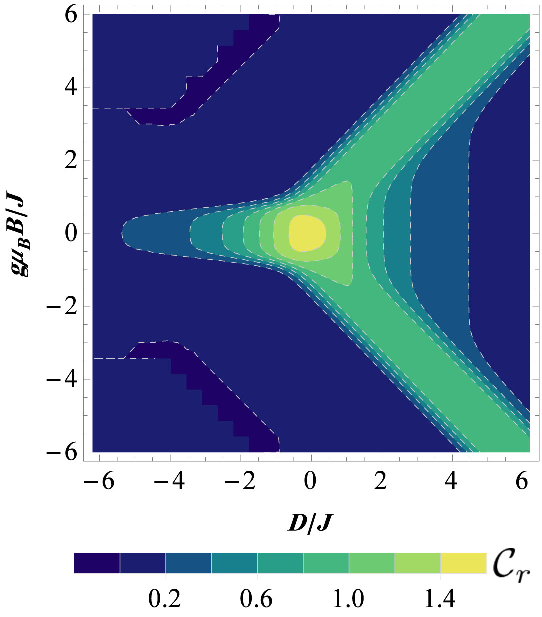} %
\includegraphics[scale=0.48,trim=00 00 00 00, clip]{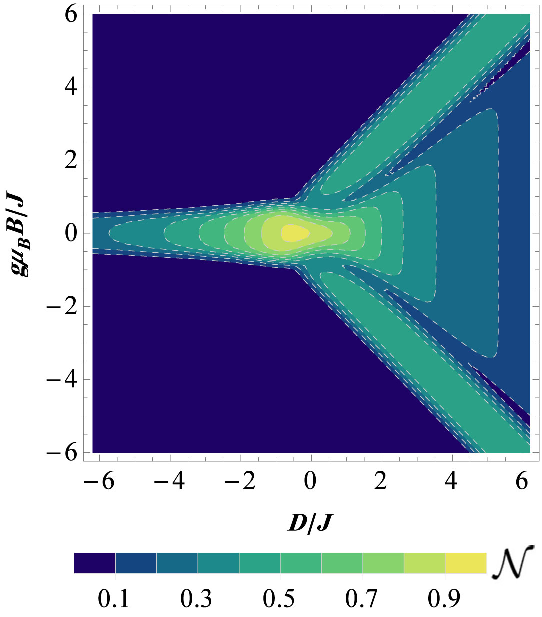} %
\includegraphics[scale=0.48,trim=00 00 00 00, clip]{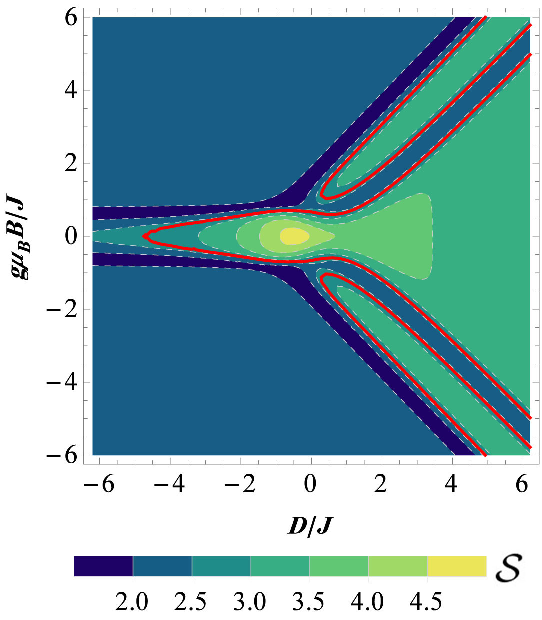}  \put(-491,130){
{\white\textbf{(a)}}} \put(-360,130){{\white\textbf{(b)}}} \put(-232,130){{\white\textbf{(c)}}}
\put(-103,130){{\white\textbf{(d)}}}
\par
\includegraphics[scale=0.48,trim=00 00 00 00, clip]{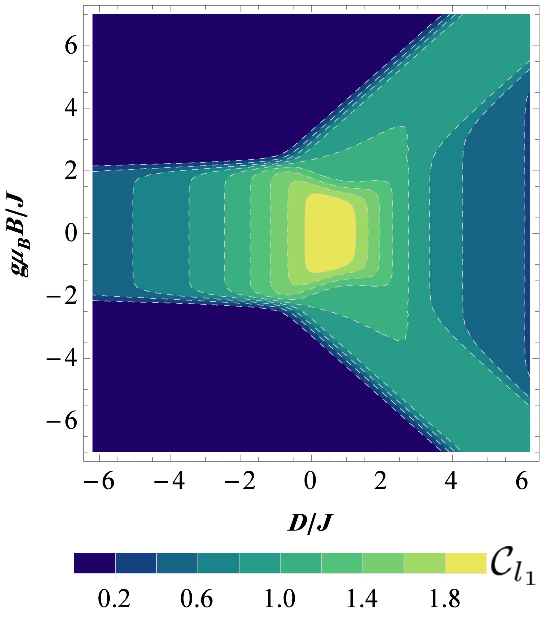} %
\includegraphics[scale=0.48,trim=00 00 00 00, clip]{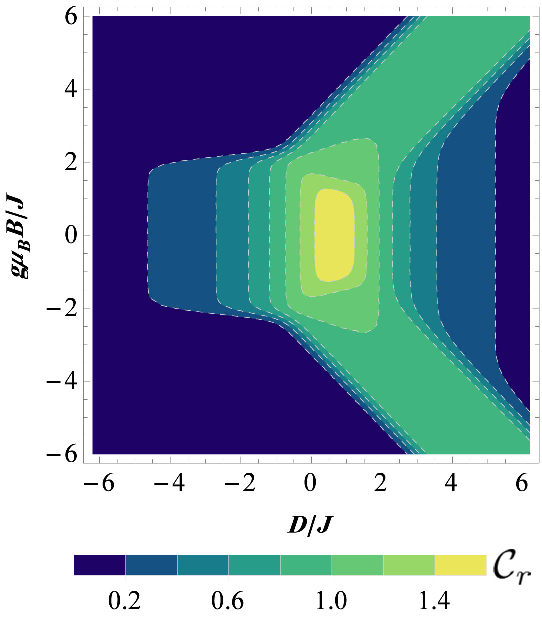} %
\includegraphics[scale=0.48,trim=00 00 00 00, clip]{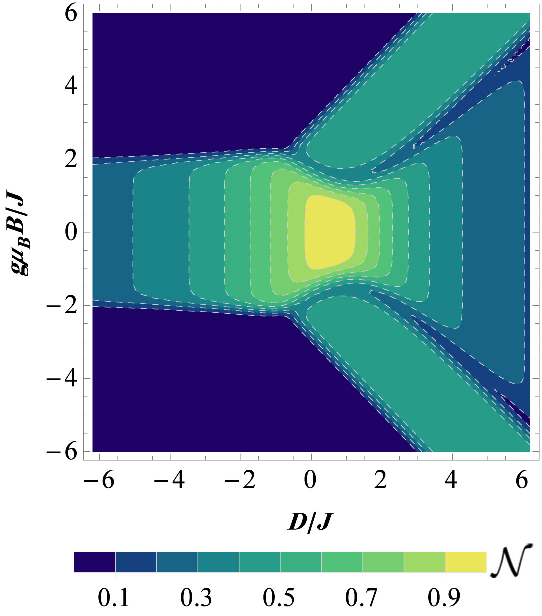} %
\includegraphics[scale=0.48,trim=00 00 00 00, clip]{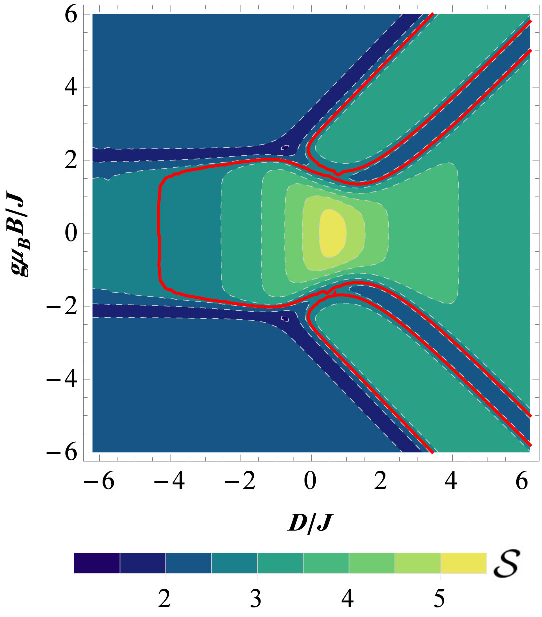} \put(-491,130){
{\white\textbf{(e)}}} \put(-360,130){{\white\textbf{(f)}}} \put(-232,130){{\white\textbf{(g)}}}
\put(-103,130){{\white\textbf{(h)}}}
\par
\includegraphics[scale=0.48,trim=00 00 00 00, clip]{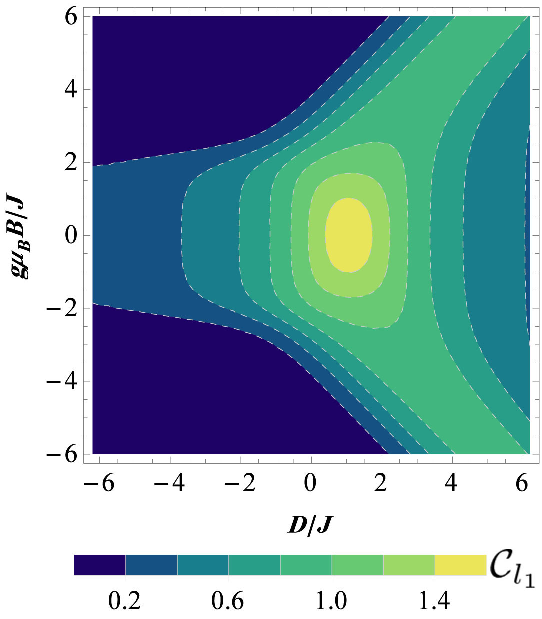} %
\includegraphics[scale=0.48,trim=00 00 00 00, clip]{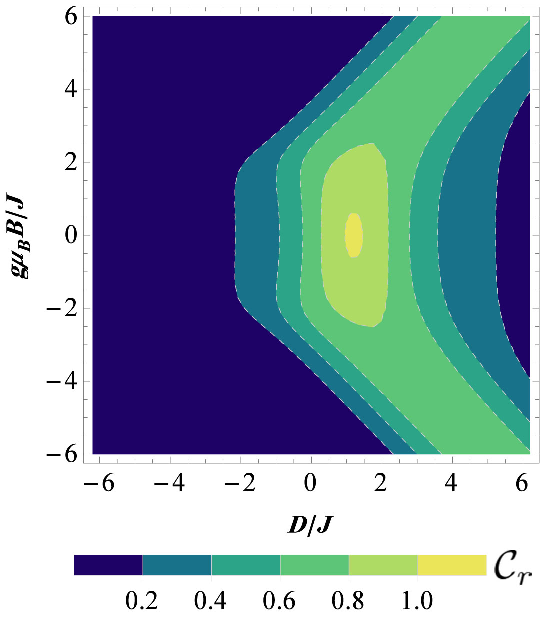} %
\includegraphics[scale=0.48,trim=00 00 00 00, clip]{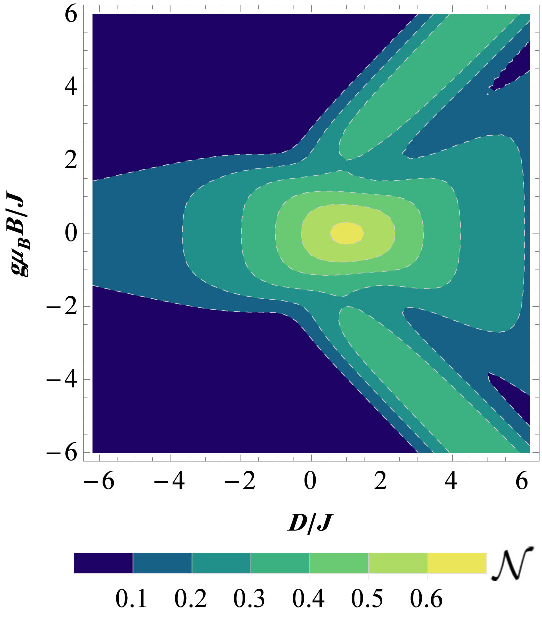} %
\includegraphics[scale=0.48,trim=00 00 00 00, clip]{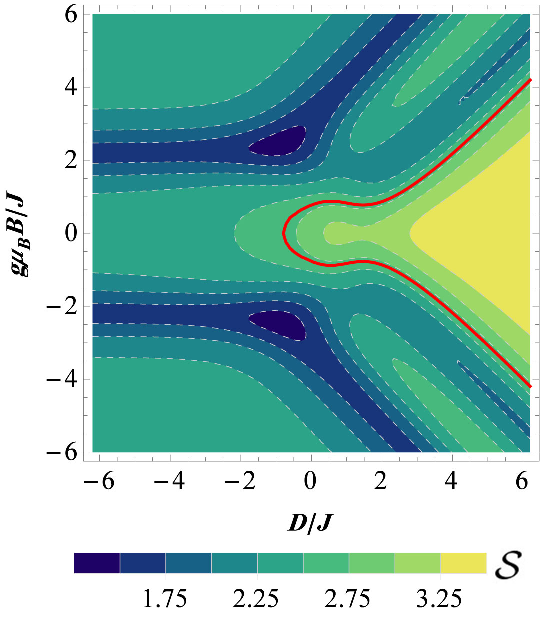} \put(-491,130){
{\white\textbf{(i)}}} \put(-360,130){{\white\textbf{(j)}}} \put(-232,130){{\white\textbf{(k)}}}
\put(-103,130){{\white\textbf{(l)}}}
\caption{ Upper panels: Density plots in the $g\mu _{B}B/J-D/J$
plane of: (a) $l_1 $-norm of quantum coherence $\mathcal{C}_{l_{1}}$, (b)
 relative entropy of coherence $\mathcal{C}_{r}$, (c) negativity $%
\mathcal{N}$, and (d) EU-steering $\mathcal{S}$, where $k_{B}T/J=0.2 $ and $%
\Delta/J=0.5$. Middle panels: Density plots of the functions with $k_{B}T/J=0.2 $ and $\Delta/J=2 $. Lower
panels: Density plots of the functions with $k_{B}T/J=0.6 $ and $\Delta/J=2$. { The red lines in graphs (d), (h), and (l) represent the lower bound of EU-steering \eqref{eusteering}, i.e. $\mathcal{S}=8/3$.}}
\label{Fig.1.1}
\end{figure*}

Figure \ref{Fig.1.1} displays the $l_{1}$-norm of quantum coherence $\mathcal{
C}_{l_{1}}$,
relative entropy of coherence $\mathcal{
C}_{r}$, negativity $\mathcal{
N}$, and the EU-steering $\mathcal{S}$ of the
considered system (\ref{gho}) as
functions of the magnetic field and the uniaxial single-ion anisotropy.
Various values of the temperature $k_{B}T/J$ and exchange anisotropy
parameter $\Delta/J $ are assumed. The density plots of the four functions are
displayed in the upper panels, where $k_{B}T/J=0.2$ and $\Delta/J =0.5$. In
Fig. \ref{Fig.1.1}(a), the pronounced increase in coherence estimated by $\mathcal{%
C}_{l_{1}}$ is observed as magnetic field values approach zero, suggesting
an intriguing interplay between intrinsic coherence and the system's
response to external magnetic fields.  This behavior can be attributed to the alignment of spin states in the absence of a magnetic field, enhancing overall coherence. Moreover, $\mathcal{%
C}_{l_{1}}$ manifests symmetrically behavior around the $g\mu
_{B}B/J$-axis, exhibiting identical characteristics regardless of $g\mu
_{B}B/J$ being positive or negative.  On one hand, although the negative $D/J$ values have a much less pronounced effect than the magnetic field, clearly the numerical value of $C_{l_1}$ changes by a factor of two-three in the range plotted.
On the other hand,  a positive shift in this single-ion anisotropy alters the coherence feature more dramatically, indicating a non-linear
interplay between these parameters. Note that some asymmetry in $\operatorname{sgn}(D)$ is expected from the Hamiltonian, as it is a switch between easy-axis and easy-plane behavior \cite{Ap2}.
Importantly, the maximum coherence area
resides within the convergence of weak magnetic fields and correspondingly
weak uniaxial single-ion anisotropy coupling.

The relative coherence in Fig. \ref{Fig.1.1}(b) reveals a resemblance to the behavior depicted in Fig. \ref{Fig.1.1}(a). In such circumstances, the attenuation of relative coherence
with escalating negative $D/J$ values reflects the diminishing influence of
external factors on the coherence structure. This can be understood as a
competition between the magnetic field-induced alignment and the
anisotropy-induced distortion of spin states, leading to a reduction in
coherence. The consistency of behavior across different parameter regimes
highlights the system's underlying coherence dynamics, which remain
resilient to variations in external conditions. Notably, the overall
behavior remains remarkably consistent for weak values of $D/J$ and $g\mu _{B}B/J$.

The entanglement in terms of
negativity depicted in Fig. \ref{Fig.1.1}(c) demonstrates a high degree of
concordance with $\mathcal{C}_{l_{1}}$. As the values of the coupling $|D/J|$
incessantly increase, the negativity exhibits a subtle decline at weaker $%
g\mu _{B}B/J$ values. Conversely, at higher $g\mu _{B}B/J$ values and
positive $D/J$ values, a counterintuitive phenomenon emerges, characterized
by a pronounced intensification of the entanglement.
These observations are similar to steering behavior, as illustrated by the area in Fig. \ref{Fig.1.1}(d) that exceeds the red line. This specific area represents the ideal steering behavior and resembles the entanglement pattern, albeit for elevated entanglement values

Besides, by increasing the exchange
anisotropy parameter~($\Delta/J =2.0$) at a low temperature with $%
k_{B}T/J=0.2$, an increase in the four quantum resources (depicted in the
middle panels) can be observed. These confirm the results plotted in Fig. %
\ref{Fig:1}. On the other hand, the $l_{1}$-norm of quantum coherence exemplified by  $%
\mathcal{C}_{l_{1}}$, displays a notable augmentation within the
range $g\mu _{B}B/J\in \lbrack 2,-2]$ for negative $D/J$ values.
Furthermore, this increase persists with rising $g\mu _{B}B/J$ for positive $%
D/J$ values. Despite the elevated $\Delta/J $, a gradual attenuation of
relative coherence $%
\mathcal{C}_{r}$ emerges with escalating negative $D/J$ values.

For negativity, we witness a significant expansion of the region that exhibits
maximal entanglement due to the increase in $\Delta/J $. Similarly, steering
behavior demonstrates a pronounced escalation in its maximum value and a
characteristic shift in its corresponding region. This is manifested as an
amplification of $\mathcal{S}>5$ and a displacement of its peak region
towards increasingly positive $D/J$ values.

\begin{figure*}[!t]
\centering
\includegraphics[scale=0.47,trim=00 00 00 00, clip]{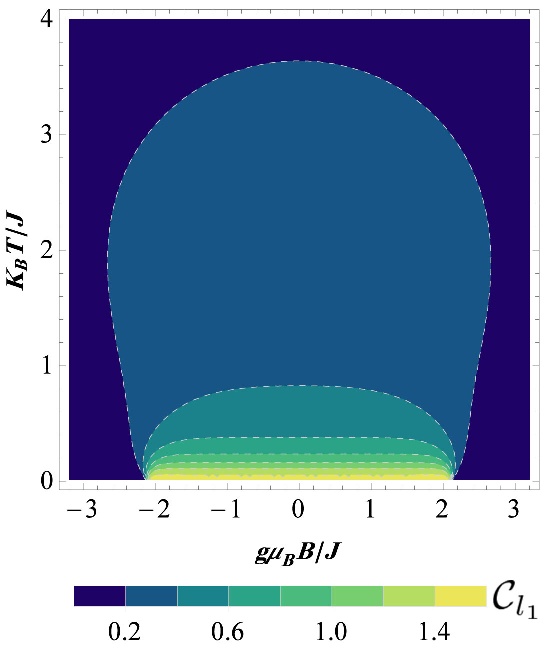} %
\includegraphics[scale=0.48,trim=00 00 00 00, clip]{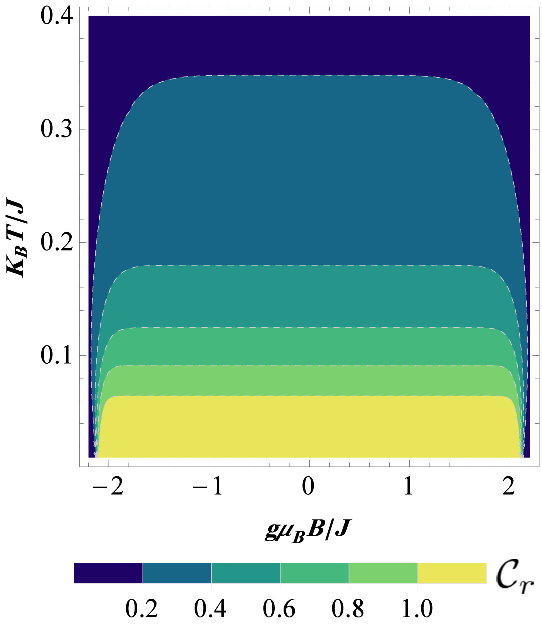} %
\includegraphics[scale=0.48,trim=00 00 00 00, clip]{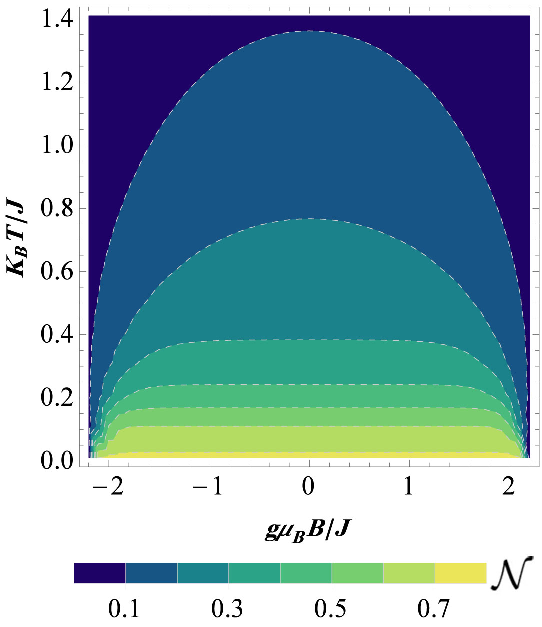} %
\includegraphics[scale=0.48,trim=00 00 00 00, clip]{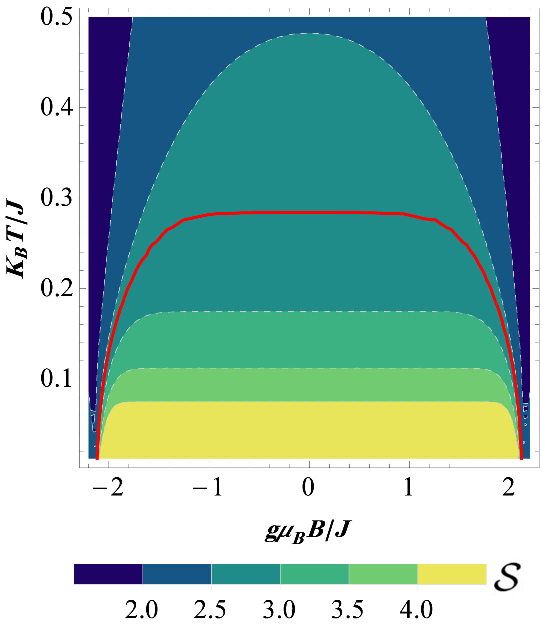} \put(-493,133){
{\white\textbf{(a)}}} \put(-360,133){{\white\textbf{(b)}}} \put(-232,133){{\white\textbf{(c)}}}
\put(-103,133){{\white\textbf{(d)}}}
\par
\includegraphics[scale=0.47,trim=00 00 00 00, clip]{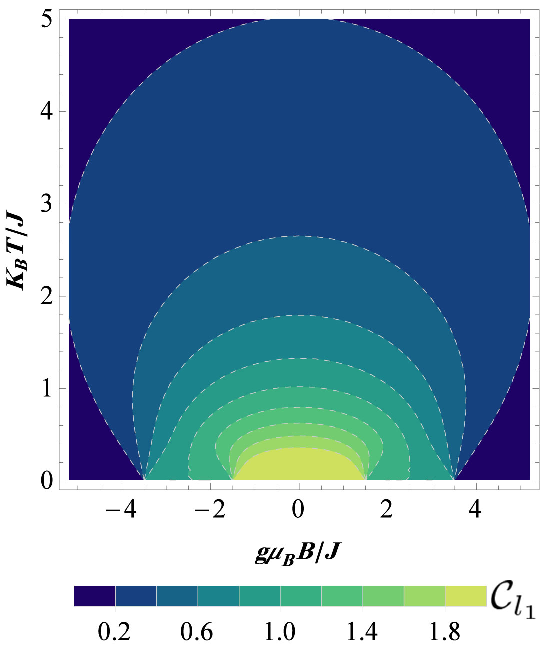} %
\includegraphics[scale=0.48,trim=00 00 00 00, clip]{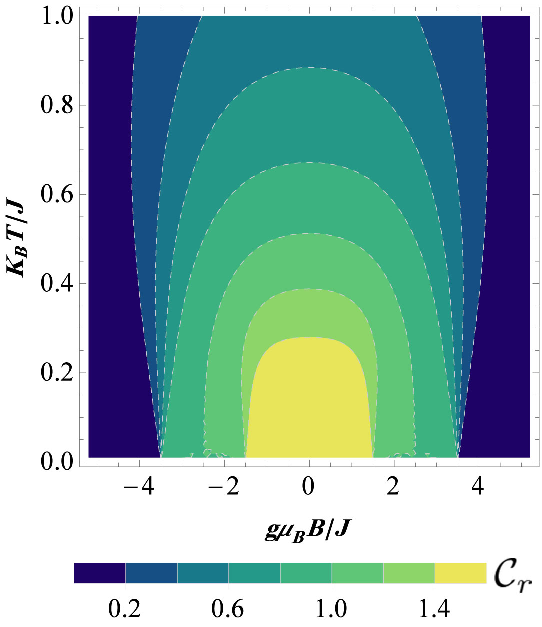} %
\includegraphics[scale=0.48,trim=00 00 00 00, clip]{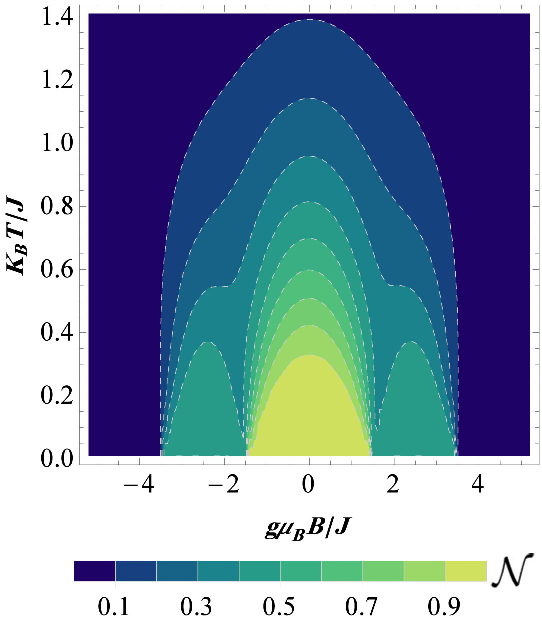} %
\includegraphics[scale=0.48,trim=00 00 00 00, clip]{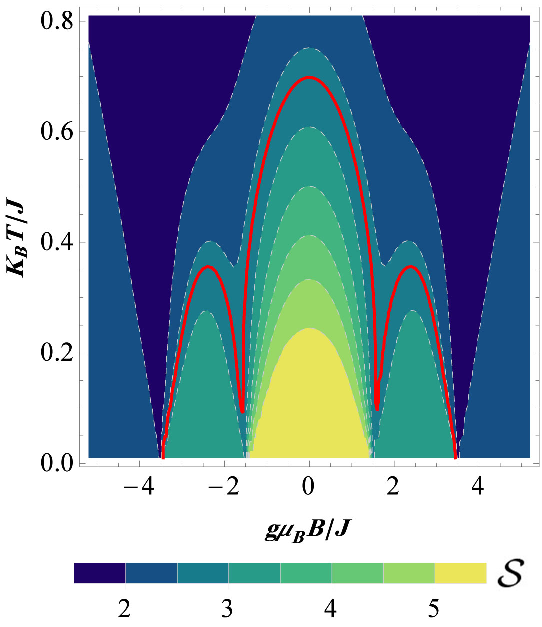} \put(-493,133){
{\white\textbf{(e)}}} \put(-360,133){{\white\textbf{(f)}}} \put(-232,133){{\white\textbf{(g)}}}
\put(-103,133){{\white\textbf{(h)}}}
\par
\includegraphics[scale=0.47,trim=00 00 00 00, clip]{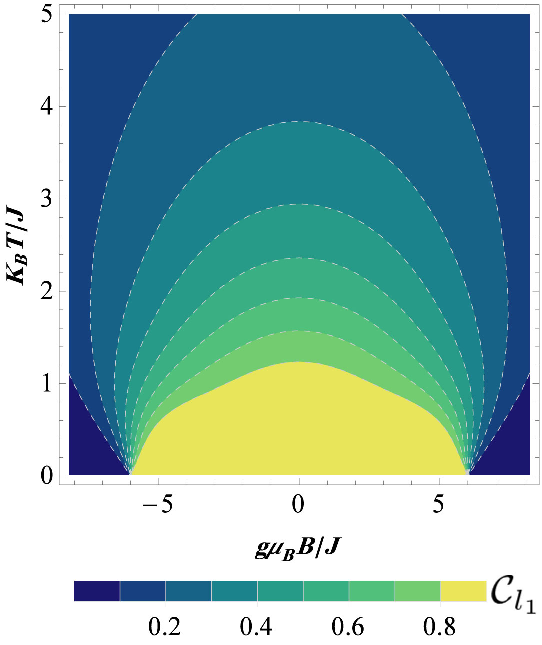} %
\includegraphics[scale=0.48,trim=00 00 00 00, clip]{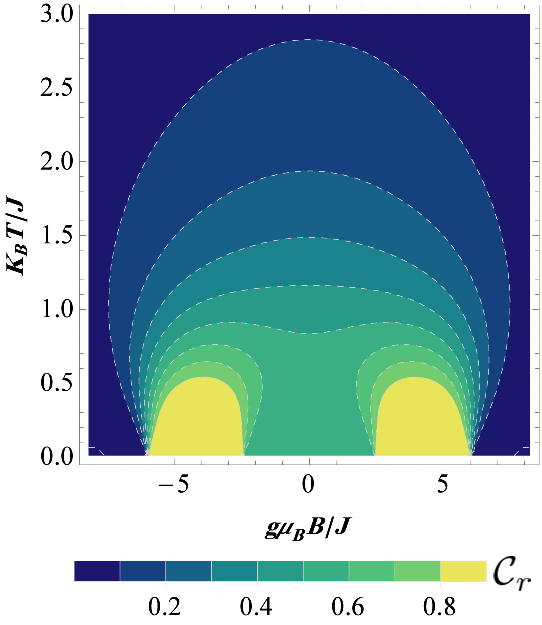} %
\includegraphics[scale=0.48,trim=00 00 00 00, clip]{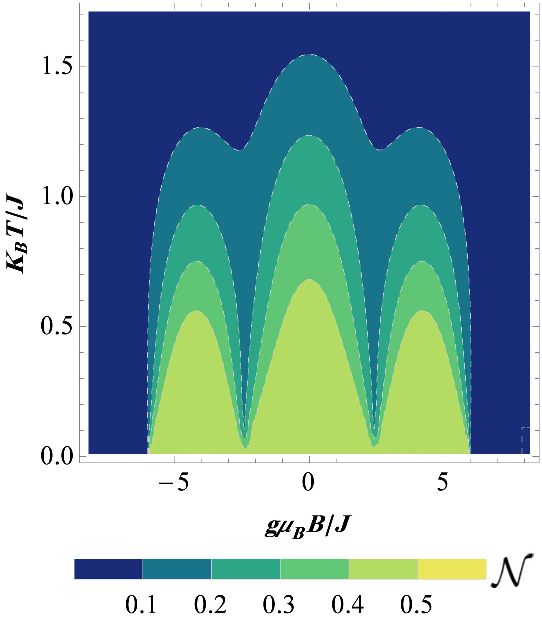} %
\includegraphics[scale=0.48,trim=00 00 00 00, clip]{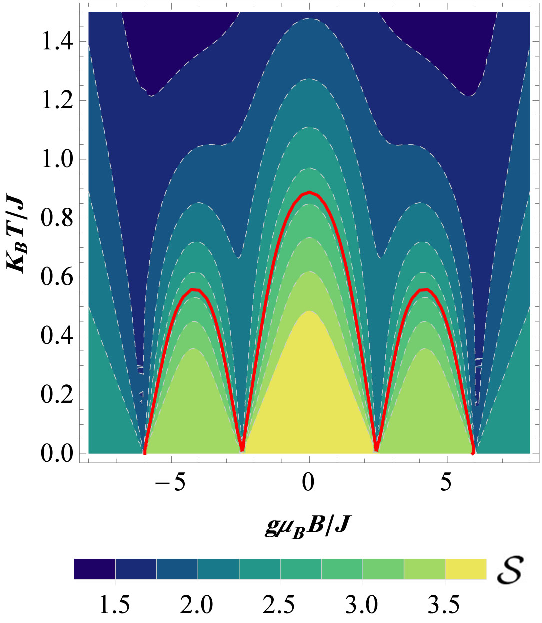} \put(-493,133){
{\white\textbf{(i)}}} \put(-360,133){{\white\textbf{(j)}}} \put(-232,133){{\white\textbf{(k)}}}
\put(-103,133){{\white\textbf{(l)}}}
\caption{ Upper panels: Density plots in the $k_{B}T/J-g\mu _{B}B/J$
plane of:  (a) $l_{1}$-norm of quantum coherence $\mathcal{C}_{l_{1}}$,
(b) the relative entropy of coherence $\mathcal{C}_{r}$, (c) negativity $%
\mathcal{N}$, and (d) EU-steering $\mathcal{S}$, where $D/J=-3$ and $\Delta/J
=2$. Middle panels: Density plots of the functions with $D/J=0.5$ and $\Delta/J =2$. Lower panels: Density
plots of the functions with $D/J=3$ and $\Delta/J =2$. { The red lines illustrate the lower bound of EU-steering  $\mathcal{S}=8/3$.}}
\label{Fig.2}
\end{figure*}

When we elevate the temperature to the value $k_{B}T/J=0.6$ while holding $%
\Delta/J $ at $2.0$, a decline in all four quantum correlation metrics is
depicted in the lower panels. The $l_{1}$-norm of  coherence, measured by $\mathcal{C}_{l_{1}}$,
diminishes significantly, reaching a maximum value at a lower $1.5$.
Moreover, relative coherence $\mathcal{C}_{r}$ entirely disappears for negative $D/J$ values
below $-2.5$, as evidenced by $\mathcal{C}_{r}=0$.

The same scenario is repeated for negativity as well so that the maximally entangled state transforms into the partially entangled state at weaker $D/J$ and $g\mu
_{B}B/J$ values.  Steering behavior becomes restricted
to positive $D/J$ values. While the region showcasing steering increases
with rising $g\mu _{B}B/J$, the overall trend signifies a weakening towards
disappearance.

These observations show how increasing temperature adversely affects quantum correlations. The system becomes more susceptible to classical behaviors, with only certain correlations exhibiting resilience even at elevated temperatures.
Thus, to understand more details about the effects of temperature on the investigated functions, we illustrate Fig. \ref{Fig.2}. This figure discusses the influence of varying $D/J$ values on the four
key quantum correlation metrics within the $(g\mu _{B}B/J-k_{B}T/J)$-plane
at a fixed value of the exchange anisotropy coupling $\Delta/J =2.0$. In the
upper panels of Fig. \ref{Fig.2}, where $D/J$ assumes a negative value, i.e.
$D/J=-3.0$, we observe a uniform behavior across all four quantifiers,
mirroring the findings for negative $D/J$ in Fig. \ref{Fig.1.1}. This
behavior entails a peak in the quantum correlations as temperatures approach
zero and $g\mu _{B}B/J$ remains low. As temperature and magnetic field $g\mu
_{B}B/J$ increase, a gradual decline in these correlations ensues,
ultimately leading to their vanishing.

Notably, $\mathcal{C}%
_{l_{1}}$ stands out with its extensive presence across the $(g\mu
_{B}B/J-k_{B}T/J)-$plane, exhibiting minimal impact from negative $D/J$. In
addition, $\mathcal{C}%
_{r}$ demonstrates a marked difference, disappearing
entirely for $k_{B}T/J\gtrsim 0.35$. For $\mathcal{N}$, we observe a gradual
progression from partial entanglement to complete separability as
temperature increases. { Steering emerges as the most susceptible correlation,
ceasing altogether for $k_{B}T/J\gtrsim 0.3$ due to $\mathcal{S}<8/3$.} This aligns with the established
hierarchical order of quantum correlations \cite{prl2007}. Based on these observations, we
can conclude that negative $D/J$ values have a substantial diminishing
effect on quantum correlations. This trend suggests a propensity towards
transforming the quantum system into a classical state (i.e. a separable state).

In the middle panels of Fig. \ref{Fig.2}, assuming $D/J$ takes weak
positive values with $D/J=0.5$, we observe a general improvement in the four
quantum quantifiers. The maximal value of $\mathcal{C}_{l_{1}}$ exceeds $1.8$, the region where $\mathcal{C}_{r}$ reaches its maximal value expands, the negativity approaches its maximum value, and also the
steering increases. A noticeable divergence in behavior arises between coherence and
the other two correlations, i.e. entanglement and steering. While coherence
maintains a consistently high presence across the $(g\mu _{B}B/J-k_{B}T/J)$%
-plane, negativity and EU-steering display a patchwork pattern, alternating
between regions of weak and strong correlations. This observation coincides
with the findings in Fig. \ref{Fig.1.1} for the positive side of $D/J$.

Increasing the value of $D/J$ to $3.0$ (bottom row of Fig. \ref{Fig.2}) results
in an expansion of the quantum correlation regions, albeit with a decline in
their peak values. As can be seen, coherence, entanglement, and steering depreciate as $%
D/J$ increases. Thereby, the increase in positive $D/J$ expands the regions exhibiting quantum correlations, but the peak correlation values diminish with increasing $D/J$.

Thus, Fig. \ref{Fig.2} demonstrates that rising temperature
propels the quantum system towards a classical system. This is evidenced by
the enhanced decoherence, disentanglement, and eventual vanishing of
steering between the two spins.  Moreover, low magnetic field $g\mu _{B}B/J$
values are crucial for preserving strong quantum correlations and achieving
maximal entanglement.

\section{Conclusion and outlook}

\label{Conclusions}
This paper conducted a comprehensive examination of quantum information
resources, encompassing the $l_{1}$-norm of quantum coherence, relative
coherence, entanglement, and steering, for the quantum spin-1 Heisenberg dimer system. We investigated
this system under the influence of magnetic field and uniaxial single-ion
anisotropy. At thermal equilibrium, we derived the final density operator of
the system and provided a concise mathematical framework elucidating the
quantum correlation metrics used in our analysis. Our findings demonstrated
that the system exhibits a maximally entangled state at absolute zero temperature for weak uniaxial single-ion anisotropy and
magnetic field values. Increasing the magnetic field transitions the system
towards a separable state while increasing the uniaxial
single-ion anisotropy leads to a partially entangled state. Notably,
enhancing the exchange anisotropy parameter significantly reinforces quantum
correlations, thereby increasing the range of the entangled state.

Considering the influence of temperature, our results revealed that
increasing temperature facilitates the system's transition towards separable
states, resulting in a reduction in coherence, entanglement, and steering.
While coherence demonstrates relative resilience to temperature increase,
the quantum resources associated with relative coherence, entanglement,
and steering exhibit greater susceptibility to thermal fluctuations. The exchange
anisotropy parameter serves as a mitigating factor against the detrimental
effects of temperature on quantum correlations, even with increased magnetic
field strength. However, under such conditions, our system undergoes a
transition from a maximally entangled state to a partially entangled state.
Furthermore, the direction of the uniaxial single-ion anisotropy parameter
significantly influences the investigated phenomena. The
generation of quantum resources exhibits a stronger dependence on the single-ion anisotropy
parameter when it is positive, and the strength of the magnetic field
further modulates this dependence.

\textbf{In conclusion}, to maximize quantum coherence, entanglement, and steering in
the spin-1 Heisenberg dimer system, it was found essential to primarily
reduce the temperature, increase the exchange anisotropy parameter, mitigate the influence of the magnetic field, and adjust the uniaxial single-ion
anisotropy parameter to a weak positive value. Besides, as the magnetic field increased, it was imperative to proportionally increase the positive
uniaxial single-ion anisotropy to maintain the system's quantumness.

The spin-1 Heisenberg dimer system offers significant potential for applications in quantum communication tasks, particularly as a resource for entanglement distribution and coherence-based protocols. The findings demonstrate that the system can maintain entanglement and coherence across various parameter values, making it suitable for controlled quantum communication environments. The system's maximally entangled state at absolute zero temperature under weak magnetic fields and uniaxial single-ion anisotropy conditions is ideal for quantum teleportation, superdense coding, and entanglement swapping. Moreover, the robustness of coherence under certain conditions suggests potential for applications in tasks requiring stable quantum superposition states, such as quantum key distribution and quantum-enhanced sensing. The exchange anisotropy parameter emerges as a vital control mechanism to counteract thermal and magnetic field effects, thereby preserving entanglement and coherence. This indicates that fine-tuning the anisotropy parameter could enable precise management of quantum resources, enhancing the feasibility of deploying such systems in realistic quantum networks. The dependence of quantum resource generation on the uniaxial single-ion anisotropy parameter and its sensitivity to its sign (positive or negative) highlight the possibility of designing systems with engineered anisotropies tailored to specific quantum tasks.

Interestingly, by adjusting the magnetic field and anisotropy parameters in tandem, the spin-1 dimer system can transition between states optimized for different quantum tasks, providing flexibility for hybrid quantum information protocols. For example, states with partial entanglement may serve as resources in noisy intermediate-scale quantum devices where absolute entanglement is not required. This adaptability suggests that such systems could be integrated into quantum repeaters, where controllable transitions between entangled and separable states play a role in efficient entanglement distribution over long distances. These aspects highlight the potential of the spin-1 Heisenberg dimer system as a versatile and tunable resource for quantum technologies, aligning the presented theoretical findings with practical quantum information science objectives.


\section*{Acknowledgments}
M.Y.A. was supported in part by the University of Chinese Academy of Sciences.

\section*{Author Contributions Statement}
F.B., M.Y.A., and S.H. wrote the main manuscript text and prepared figures. All authors reviewed the manuscript.

\section*{Disclosures}
The authors declare that they have no known competing financial interests.

\section*{Data availability}
All data generated or analysed during this study are included in this published article.

\appendix

\section{Eigenenergies and eigenvectors}
\label{Sec:appendix a}
{ The eigenenergies of the spin-1
Heisenberg dimer \eqref{H} are as follows
\begin{equation}
{E_{1}}=\Delta +2D-2h,\text{ \ \ }{E_{2}}=J+D+h,\text{ \ \ }E_{3}=-J+D+h,%
  \label{EIGENVALUES}
\end{equation}%
\begin{equation}
E_{4}=\Delta +2D+2h,\text{ \ \ } {E_{5}}=-\frac{\Delta}{2} +D+\frac{1}{2}\sqrt{(\Delta-2D)
^{2}+8J^{2}},
\end{equation}%
\begin{equation}
{E_{6}}=-\frac{\Delta}{2} +D-\frac{1}{2}\sqrt{%
(\Delta-2D)
^{2}+8J^{2}}, \text{ \ \ } {E_{7}}=-\Delta +2D,
\end{equation}%
\begin{equation}
{E_{8}}=J+D-h,\text{ \ \ }E_{9}=-J+D-h,
\end{equation}%
and the associated eigenvectors are respectively given by
\begin{equation}\label{state1.2.3}
\left\vert \psi _{1}\right\rangle =\left\vert -1,-1\right\rangle ,\text{ \ \
}\left\vert \psi _{2}\right\rangle =\frac{1}{\sqrt{2}}\left( \left\vert
0,1\right\rangle +\left\vert 1,0\right\rangle \right),
\end{equation}
\begin{equation}
\left\vert \psi _{3}\right\rangle =\frac{1}{\sqrt{2}}\left( -\left\vert
0,1\right\rangle +\left\vert 1,0\right\rangle \right) ,\text{ \ \ }
\left\vert \psi _{4}\right\rangle =\left\vert 1,1\right\rangle,
\end{equation}
\begin{equation}
\left\vert \psi _{5}\right\rangle =\frac{1}{\sqrt{2+\Lambda _{+}^{2}}}\left(
\left\vert -1,1\right\rangle +\Lambda _{+}\left\vert 0,0\right\rangle
+\left\vert 1,-1\right\rangle \right),
\end{equation}
\begin{equation}
\left\vert \psi
_{6}\right\rangle =\frac{1}{\sqrt{2+\Lambda _{-}^{2}}}\left( \left\vert
-1,1\right\rangle +\Lambda _{-}\left\vert 0,0\right\rangle +\left\vert
1,-1\right\rangle \right),
\end{equation}
\begin{equation}
\left\vert \psi _{7}\right\rangle =\frac{1}{\sqrt{2}}\left( -\left\vert
-1,1\right\rangle +\left\vert 1,-1\right\rangle \right),
\end{equation}
\begin{equation}
\left\vert \psi _{8}\right\rangle =\frac{1}{\sqrt{2}}\left( \left\vert
-1,0\right\rangle +\left\vert 0,-1\right\rangle \right),
\end{equation}
\begin{equation}
\left\vert \psi _{9}\right\rangle =\frac{1}{\sqrt{2}}\left( -\left\vert
-1,0\right\rangle +\left\vert 0,-1\right\rangle \right),
\end{equation}
where ${{\Lambda _{\pm }}}={{(\Delta -2D\pm \sqrt{(\Delta-2D)
^{2}+8J^{2}})/2J}}$. }

\section{Some notes on negativity}
\label{Sec:appendix b}
{ Negativity is generally used to detect and quantify entanglement in bipartite states. If the negativity is greater than zero, the state is entangled. Conversely, if the negativity is zero, the state is separable. However, one question arises: can the negativity be zero even if the state is entangled? The answer is yes but with some subtleties:  bound entanglement and higher dimensions.  There are entangled states, known as bound entangled states \cite{PhysRevLett.80.5239}, that have zero negativity. These states are entangled but cannot be distilled into pure entangled states using local operations and classical communication (LOCC). Bound entangled states are a special class of mixed states for which the partial transpose is positive semidefinite (PPT criterion), hence their negativity is zero, yet they are entangled. Bound entanglement typically occurs in higher-dimensional systems (e.g.,
$3\otimes3$ or higher-dimensional Hilbert spaces). In $2\otimes2$ or $2\otimes3$ systems, the Peres-Horodecki criterion states that a state is separable if and only if its partial transpose is positive semidefinite, thus implying that a non-positive partial transpose (and hence non-zero negativity) is a necessary and sufficient condition for entanglement in these lower-dimensional cases. Thus, while negativity being zero usually indicates separability, there are exceptions in the form of bound entangled states, where the state is entangled despite having zero negativity.}


\begin{thebibliography}{99}
\bibitem{yao2012scalable} N.Y. Yao, L. Jiang, A.V. Gorshkov, P.C. Maurer, G.
Giedke, J.I. Cirac, and M.D Lukin, Scalable architecture for a room temperature
solid-state quantum information processor. \textit{Nature Commun.}, \textbf{3}, 800 (2012).

\bibitem{leuenberger2001quantum} M.N. Leuenberger and D. Loss, Quantum
computing in molecular magnets. \textit{Nature} \textbf{410}, 789 (2001).

\bibitem{devoret2013superconducting} M.H. Devoret and R.J. Schoelkopf,
Superconducting circuits for quantum information: an outlook. \textit{Science}, \textbf{339}, 6124 (2013).

\bibitem{PhysRevResearch.2.013062} H.Y. Xu and Y.C. Lai, Anomalous chiral edge
states in spin-1 Dirac quantum dots. \textit{Phys. Rev. Research} \textbf{2}, 013062 (2020).

\bibitem{blatt2008entangled} R. Blatt and D. Wineland, Entangled states of
trapped atomic ions. \textit{Nature} \textbf{453}, 7198 (2008).

\bibitem{wolfowicz2021quantum} G. Wolfowicz, F.J. Heremans, C.P. Anderson,
S. Kanai, H. Seo, A. Gali, G. Galli, and D.D. Awschalom, Quantum guidelines for
solid-state spin defects. \textit{Nat. Rev. Mat.} \textbf{6}, 906 (2021).

\bibitem{PhysRevB.102.184419} H. \ifmmode \v{C}\else \v{C}\fi{}en\ifmmode
\v{c}\else \v{c}\fi{}arikov\'a, and J. Stre\ifmmode \v{c}\else \v{c}\fi{}ka,
Unconventional strengthening of the bipartite entanglement of a mixed
spin-(1/2,1) Heisenberg dimer achieved through Zeeman splitting. \textit{Phys. Rev. B} \textbf{102}, 184419 (2020).

\bibitem{PhysRevA.72.022314} I. Bose and A. Tribedi, Thermal entanglement
properties of small spin clusters. \textit{Phys. Rev. A} \textbf{72}, 022314 (2005).

\bibitem{Bonner1981} J.C. Bonner, H.W.J. Blöte,  H. Beck, and  G. Müller,  Quantum Spin Chains. In: Bernasconi, J., Schneider, T. (eds) Physics in One Dimension (Springer Series in Solid-State Sciences, vol 23, pp 115–128, Springer, Berlin, Heidelberg, 1981).

\bibitem{PhysRevB.102.064414} F. Souza, L.M. Ver\'{\i}ssimo, J. Stre\ifmmode
\v{c}\else \v{c}\fi{}ka, M.L. Lyra, and M.S.S. Pereira, Exact and density matrix
renormalization group studies of two mixed spin-$(\frac{1}{2},\frac{5}{2}, \frac{1}{2})$ branched-chain models developed for a heterotrimetallic {Fe-Mn-Cu} coordination polymer. \textit{Phys. Rev. B} \textbf{102}, 064414
(2020).

\bibitem{sugimoto2023quasi} T. Sugimoto and T. Tohyama, Quasi-fractionalization
of edge spin in chirality-assisted cluster-based {H}aldane state on triangular spin tube. \textit{Commun. Phys.} \textbf{6}, 291 (2023).

\bibitem{arian2020spin} H. Arian Zad, A. Trombettoni, and N. Ananikian, Spin-1/2
{I}sing--{H}eisenberg Cairo pentagonal model in the presence of an external
magnetic field: Effect of Land{\'e} g-factors. \textit{Europ. Phys. J. B}
\textbf{93}, 200 (2020).

\bibitem{PhysRevB.103.184415} T. Verkholyak and J. Stre\ifmmode \v{c}\else
\v{c}\fi{}ka, Modified strong-coupling treatment of a spin-$\frac{1}{2}$ {H}eisenberg trimerized chain developed from the exactly solved {I}sing-{H}
eisenberg diamond chain. \textit{Phys. Rev. B} \textbf{103}, 184415 (2021).

\bibitem{hida2020ground} K. Hida, Ground-State Phases of Alternating-Bond $S=1$ Diamond Chains. \textit{J. Phys. Soc. Jap.} \textbf{89}, 2 (2020).

\bibitem{PhysRevB.101.195110} D. Dey, K. Sambunath, and S. Ramasesha,
Magnetization plateaus of spin-$\frac{1}{2}$ system on a $5/7$ skewed
ladder. \textit{Phys. Rev. B} \textbf{101}, 195110 (2020).

\bibitem{PhysRevA.101.033607} L. Zhou and Q. Du, Floquet topological phases
with fourfold-degenerate edge modes in a driven spin-1/2 Creutz ladder.
\textit{Phys. Rev. B} \textbf{101}, 033607 (2020).

\bibitem{PhysRevB.84.054451} G. De Chiara, M. Lewenstein, and A. Sanpera,
Bilinear-biquadratic spin-1 chain undergoing quadratic Zeeman effect.
\textit{Phys. Rev. B} \textbf{84}, 054451 (2011).

\bibitem{kruk2001nuclear} D. Kruk, T. Nilsson, and J. Kowalewski, Nuclear spin
relaxation in paramagnetic systems with zero-field splitting and arbitrary
electron spin. \textit{Phys. Chem. Chem. Phys.} \textbf{3}, 4907 (2001).

\bibitem{benabdallah2020quantum} F. Benabdallah, A. Slaoui, and M. Daoud,
Quantum discord based on linear entropy and thermal negativity of
qutrit--qubit mixed spin chain under the influence of external magnetic
field. \textit{Quantum Inf. Process.} \textbf{19}, 252 (2020).

\bibitem{benabdallah2021quantum} F. Benabdallah and M. Daoud, Dynamics of
quantum discord based on linear entropy and negativity of qutrit-qubit
system under classical dephasing environments. \textit{Eur. Phys. J. D.}
\textbf{75}, 3 (2021).

\bibitem{khalil2022robustness} E.M. Khalil and M.Y. Abd-Rabbou, Robustness of
a teleported state influenced by dipole interaction and magnetic field under
intrinsic decoherence. \textit{Optik} \textbf{267}, 169703 (2022).

\bibitem{PhysRevE.106.034122} F. Benabdallah, A.U. Rahman, S. Haddadi and
M. Daoud, Long-time protection of thermal correlations in a hybrid-spin system under random telegraph noise. \textit{Phys. Rev. E} \textbf{106}, 034122 (2022).

\bibitem{PhysRevE.106.0341221} F. Benabdallah, H. Arian Zad. M. Daoud and N.
Ananikian, Dynamics of quantum correlations in a qubit-qutrit spin system
under random telegraph noise. \textit{Phys. Scr.} \textbf{96}, 125116 (2021).

{ \bibitem{Yurischev2022} A.V. Fedorova and M.A. Yurischev, Behavior of quantum discord, local quantum uncertainty, and local quantum Fisher information in two-spin-1/2 Heisenberg chain with DM and KSEA interactions. \textit{Quantum Inf. Process.} \textbf{21},  92 (2022).}


\bibitem{RevModPhys.81.865} R. Horodecki, P. Horodecki, M. Horodecki, and K.
Horodecki, Quantum entanglement. \textit{Rev. Mod. Phys.} \textbf{81}, 865 (2009).

\bibitem{RevModPhys.92.015001} A. Streltsov, G. Adesso, and M.B.
Plenio, Colloquium: Quantum coherence as a resource. \textit{Rev. Mod. Phys.} \textbf{89}, 041003 (2017).

\bibitem{RevModPhys.89.041003} R. Uola, A.C.S. Costa, H. Chau Nguyen, and O. Gühne, Quantum steering. \textit{Rev. Mod. Phys.} \textbf{92}, 015001 (2020).

\bibitem{artur2022} A. Czerwinski, Quantum tomography of entangled qubits by time-resolved single-photon counting with time-continuous measurements. \textit{Quantum Inf. Process.} \textbf{21}, 332 (2022).


\bibitem{nielsen2010quantum} M.A. Nielsen and I.L. Chuang, Quantum
computation and quantum information (Cambridge University Press, 2010).

\bibitem{asadali2024} A. Ali, S. Al-Kuwari, and S. Haddadi, Trade-off relations of quantum resource theory in Heisenberg models. \textit{Phys. Scr.} \textbf{99}, 055111 (2024).

\bibitem{asadali20242} A. Ali, S. Al‑Kuwari, M. T. Rahim, M. Ghominejad, H. Ali, and S. Haddadi, A study on thermal quantum resources and probabilistic teleportation in spin‑1/2 Heisenberg XYZ+DM+KSEA model under variable Zeeman
splitting. \textit{Appl. Phys. B} \textbf{130},  177 (2024).


\bibitem{schrodinger1935mathematical} E. Schr{\"o}dinger, Mathematical
proceedings of the cambridge philosophical society. \textit{Math Pro Camb
Philos Soc.} \textbf{31}, 555 (1935).

\bibitem{PhysRevLett.80.5239} M. Horodecki, P. Horodecki, and R. Horodecki, Mixed-state entanglement and distillation: Is there a ``bound'' entanglement
in nature?. \textit{Phys. Rev. Lett.} \textbf{80}, 5239 (1998).

\bibitem{PhysRevLett.91.207901} S. Bose, Quantum Communication through an
Unmodulated Spin Chain. \textit{Phys. Rev. Lett.} \textbf{91}, 207901 (2003).

\bibitem{mehring1999concepts} M. Mehring, Concepts of spin quantum
computing. \textit{Appl. Magn. Res.} \textbf{17}, 141 (1999).

\bibitem{obada2024} A.-S. F. Obada, M. Y. Abd-Rabbou, and S. Haddadi, Does conditional entropy squeezing indicate normalized entropic uncertainty relation steering?. \textit{Quantum Inf. Process.} \textbf{23}, 90 (2024).

\bibitem{PhysRevA.94.062120} C.Y. Hsieh, Y.C. Liang, and R.K. Lee, Quantum
steerability: characterization, quantification, superactivation, and
unbounded amplification. \textit{\ Phys. Rev. A} \textbf{94}, 062120 (2016).

\bibitem{PhysRevA.99.012302} C.Y. Huang, N. Lambert, C.M. Li, Y.T. Lu, and
F. Nori, Securing quantum networking tasks with multipartite
Einstein-Podolsky-Rosen steering. \textit{Phys. Rev. A} \textbf{99}, 012302 (2019).

\bibitem{PhysRevLett.70.1895} C.H. Bennett, G. Brassard, C. Gilles, J. Claude, R. Jozsa, A. Peres, and W.K. Wootters: Teleporting an unknown quantum
state via dual classical and Einstein-Podolsky-Rosen channels. \textit{Phys.
Rev. Lett.} \textbf{70}, 1895 (1993).

\bibitem{F1} F. Benabdallah, K. Elanouz, and M. Daoud, Toward the
relationship between local quantum Fisher information and local quantum
uncertainty in the presence of intrinsic decoherence. \textit{Eur. Phys. J.
Plus} \textbf{137}, 548 (2022).

\bibitem{F2} F. Benabdallah, S. Haddadi, H. Arian Zad, M.R. Pourkarimi,
M. Daoud, and N. Ananikian, Pairwise quantum criteria and teleportation in a
spin square complex. \textit{Sci. Rep.} \textbf{12}, 6406 (2022).

\bibitem{PhysRevLett.125.020404} S. Wollmann, R. Uola, and A.C.S. Costa,
Experimental demonstration of robust quantum steering. \textit{Phys. Rev.
Lett.} \textbf{125}, 020404 (2020).

\bibitem{PhysRevA.98.052351} L.F. Qiao, A. Streltsov, J. Gao, S. Rana, R.J.
Ren, Z.Q. Jiao, C.Q. Hu, X.Y. Xu, C.Y. Wang, H. Tang, A.L. Yang, Z.H. Ma, M.
Lewenstein, and X. Min Jin, Entanglement activation from quantum coherence
and superposition. \textit{Phys. Rev. A} \textbf{98}, 052351 (2018).

\bibitem{RevModPhys.89.035002} C.L. Degen, F. Reinhard, and P. Cappellaro,
Quantum sensing. \textit{Rev. Mod. Phys.} \textbf{89}, 035002 (2017).

\bibitem{PhysRevA.98.032101} D.P. Pires, I.A. Silva, E.R. deAzevedo, D.O.
Soares-Pinto, and J.G. Filgueiras, Coherence orders, decoherence, and quantum
metrology. \textit{Phys. Rev. A.} \textbf{98}, 032101 (2018).

\bibitem{ruster2016long} T. Ruster, C.T. Schmiegelow, H. Kaufmann, C.
Warschburger, F. Schmidt-Kaler, and U.G. Poschinger: A long-lived Zeeman
trapped-ion qubit. \textit{Appl. Phys. B} \textbf{122}, 254 (2016).

\bibitem{F3} F. Benabdallah, K. El Anouz, J. Strecka and M. Daoud: Thermal
non-classical correlation via skew information, quantum Fisher information,
and quantum teleportation of a spin-1/2 Heisenberg trimer system. \textit{%
Eur. Phys. J. Plus} \textbf{137}, 1096 (2022).

\bibitem{abd2022enhancing} M.Y. Abd-Rabbou, S.I. Ali, and M.M.A. Ahmed,
Enhancing the information of nonlinear {SU} (1, 1) quantum systems
interacting with a two-level atom. \textit{Opt. Quantum Elect.} \textbf{54},
9 (2022).

\bibitem{stockill2016quantum} R. Stockill, C. Le Gall, C. Matthiesen, L.
Huthmacher, E. Clarke, M. Hugues, M. Atat{\"u}re: Quantum dot spin coherence
governed by a strained nuclear environment. \textit{Nat. Commun.} \textbf{7},
12745 (2016).

\bibitem{NN1} X. Wang, H.B. Li, Z. Sun, Y.Q. Li: Entanglement in spin-1
Heisenberg chains. \textit{J. Phys. A Math. Gen.} \textbf{38}, 8703 (2005).

\bibitem{NN2} X. Wang, S.-J. Gu: Negativity, entanglement witness and
quantum phase transition in spin-1 Heisenberg chains. \textit{J. Phys. A
Math. Gen.} \textbf{40}, 10759 (2007).

\bibitem{NNN2} P. Xu, Y.-H. Hu, X.-W. Hou, Thermal quantum coherence and
correlations in a spin-1 Heisenberg model. \textit{Physica A} \textbf{491}, 282 (2018).

\bibitem{NN3} E. Albayrak, Thermal entanglement in two-qutrit spin-1
anisotropic Heisenberg model with inhomogeneous magnetic field. \textit{%
Chin. Phys. B} \textbf{19}, 090319 (2010).

\bibitem{NN5} E. Albayrak, Thermal entanglement in the XYZ model for a
two-qutrit system. \textit{Opt. Commun.} \textbf{284}, 1631 (2011).

{

\bibitem{added1} A. L. Malvezzi, G. Karpat, B. Çakmak, F. F. Fanchini, T. Debarba, and R. O. Vianna, Quantum correlations and coherence in spin-1 Heisenberg chains. \textit{Phys. Rev. B} \textbf{93}, 184428 (2016).

\bibitem{added2} J. Lambert and E. S. Sørensen, Estimates of the quantum Fisher information in the $S=1$ antiferromagnetic Heisenberg spin chain with uniaxial anisotropy. \textit{Phys. Rev. B} \textbf{99}, 045117 (2019).

\bibitem{added3} F. Dell'Anna, S. Pradhan, C. Degli E. Boschi, and E. Ercolessi, Quantum Fisher information and multipartite entanglement in spin-1 chains. \textit{Phys. Rev. B} \textbf{108}, 144414 (2023).

\bibitem{Escuer1994} A. Escuer, R. Vicente, J. Ribas, J. Jaud, B. Raynaud,  Octahedral $\mu$-oxalato-nickel(II) dinuclear complexes with water and
tridentate amines as blocking ligands: Magnetostructural correlations. \textit{Inorg. Chim. Acta} \textbf{216}, 139  (1994).}

\bibitem{Ap1} L. Stre{\v{c}}ka, M. Ja{\v{s}}{\v{c}}ur, M. Hagiwara, Y.
Narumi, J. Kuch{\'a}r, S. Kimura, K. Kindo, Magnetic behavior of a spin-1
dimer: model system for homodinuclear nickel (II) complexes. \textit{J.
Phys. Chem. Sol.} \textbf{66}, 1828 (2005).

\bibitem{Ap2} A. Ghannadan, and J Stre{\v{c}}ka, Magnetic-field-orientation
dependent thermal entanglement of a spin-1 Heisenberg dimer: The case study
of dinuclear nickel complex with an uniaxial single-ion anisotropy. \textit{%
Molecules} \textbf{26}, 3420 (2021).

{ \bibitem{Baumgratz2014} T. Baumgratz, M. Cramer, and M. B. Plenio, Quantifying coherence. \textit{Phys. Rev. Lett.} \textbf{113}, 140401  (2014).}

{ \bibitem{Hu2018} M.-L. Hu, X. Hu, J. Wang, Y. Peng, Y.-R. Zhang, H. Fan, Quantum coherence and geometric quantum discord. \textit{Phys. Rep.}  \textbf{762-764}, 1-100 (2018).}

\bibitem{COH} K. Bu, U. Singh, S.M. Fei, A.K. Pati, and J. Wu: Maximum
Relative Entropy of Coherence: An Operational Coherence Measure. \textit{%
Phys. Rev. A} \textbf{119}, 150405 (2017).

\bibitem{Peres} A. Peres, Magnetic-Field-Orientation Dependent Thermal
Entanglement of a Spin-1 Heisenberg Dimer: Separability Criterion for
Density Matrices. \textit{Phys. Rev. Lett.} \textbf{77}, 1413 (1996).

{ \bibitem{Vidal} G. Vidal, and R. F. Werner: Computable measure of entanglement. \textit{Phys. Rev. A.} \textbf{65},
032314 (2002).}

\bibitem{PhysRevA.79.022104} S. Wu, S. Yu, K. M\o {}lmer: Entropic
uncertainty relation for mutually unbiased bases. \textit{Phys. Rev. A}
\textbf{79}, 022104 (2009).

\bibitem{S1} W.-Y. Sun, D. Wang, J.-D. Shi, and L. Ye, Exploration
quantum steering, nonlocality and entanglement of two-qubit X-state in
structured reservoirs. \textit{Sci. Rep.}, \textbf{7}, 39651 (2017).

\bibitem{S2} M. Y. Abd-Rabbou, N. Metwally, M. M. A. Ahmed, and A.-S. F.
Obada, Improving the bidirectional steerability between two accelerated
partners via filtering process. \textit{Mod. Phys. Lett. A} \textbf{37}, 2250143 (2022).

\bibitem{S3} A. U. Rahman, M. Shamirzaie, and M. Y. Abd-Rabbou,
Bidirectional steering, entanglement and coherence of accelerated
qubit-qutrit system with a stochastic noise. \textit{Optik} \textbf{274},
170543 (2023).

\bibitem{costa2018entropic} A.C.S Costa, R. Uola, Roope, and O. G{\"u}hne, Entropic steering criteria:
applications to bipartite and tripartite systems. \textit{Entropy} \textbf{20}, 763 (2018).

\bibitem{prl2007}
H. M. Wiseman, S. J. Jones, and A. C. Doherty, Steering, entanglement, nonlocality, and the Einstein-Podolsky-Rosen paradox. Phys. Rev. Lett. \textbf{98}, 140402 (2007).

\end{thebibliography}
\end{document}